\documentclass[sigconf, nonacm]{acmart}
\AtBeginDocument{%
  }

\usepackage{graphicx}
\usepackage{algorithm}
\usepackage{algpseudocodex}
\usepackage{subcaption}
\usepackage{float}
\usepackage{tikz}
\usetikzlibrary{calc,shadows.blur,positioning,shapes.geometric}
\usepackage{standalone}
\usepackage{listings}
\usepackage{stfloats}
\usepackage{pgfplots}
\pgfplotsset{compat=1.18}
\usepgfplotslibrary{colormaps}

\author{Abinand Nallathambi}
\affiliation{%
  \institution{Purdue University}
  \city{West Lafayette}
  \state{IN}
  \country{USA}
}
\author{Christopher Knight}
\affiliation{%
  \institution{Argonne National Laboratory}
  \city{Argonne}
  \state{IL}
  \country{USA}
}
\author{Shantanu Ganguly}
\affiliation{%
  \institution{Unaffiliated}
  \city{}
  \state{}
  \country{}
}

\author{Wilfried Haensch}
\affiliation{%
  \institution{University of Chicago}
  \city{Chicago}
  \state{IL}
  \country{USA}
}

\author{Anand Raghunathan}
\affiliation{%
  \institution{Purdue University}
  \city{West Lafayette}
  \state{IN}
  \country{USA}
}

\renewcommand{\shortauthors}{Nallathambi et al.}

\begin{document}

\title{A3D: Agentic AI flow for autonomous Accelerator Design}



\begin{abstract}
Accelerating applications through the design of hardware accelerators can significantly enhance system performance and energy efficiency. Despite advances, such as high-level synthesis (HLS), designing accelerators for complex applications still remains highly labor-intensive, demanding considerable expertise in understanding workloads to be accelerated, hardware design, micro-architecture, and EDA tool usage, posing challenges for application domain experts. Therefore, most accelerator solutions are limited to applications with a regular predictable dataflow. Advances in AI have enabled agents that perform autonomous planning, reasoning, execution and reflection, leading to unprecedented potential for automation through agentic AI. We present A3D, an Agentic AI flow for end-to-end Automation of hardware Accelerator Design. A3D automates workload analysis, performance bottleneck identification, code refactoring for HLS compatibility and micro-architecture generation. A3D also generates diverse accelerator designs by automatically exploring the speed-area tradeoff space. Recent efforts have explored the use of AI for specific tasks such as design space exploration in HLS, leaving several tasks to still be performed manually. A3D addresses the challenges in applying modern LLMs to accelerator design by judiciously partitioning tasks among specialist agents, orchestrating process loops with specialist and verifier agents, utilizing pre-existing and custom tools, and employing agentic RAG for codebase and proprietary EDA tool documentation exploration. Our implementation of A3D, using commercial components like Claude Sonnet 4.5 and the Catapult HLS tool, demonstrates its effectiveness by generating accelerator designs with no human intervention from complex scientific applications like LAMMPS (molecular dynamics simulation) and QMCPACK (quantum chemistry).
\end{abstract}

\ccsdesc[500]{Software~Hardware tools for EDA}
\keywords{AI agents, EDA, Accelerator Design, High-Level Synthesis, Scientific computing}

\maketitle

\section{Introduction}
\label{sec:intro}

Specialization, or the design and use of hardware accelerators, is a fundamental technique in computing system design and has the potential to speed up the execution of applications by orders of magnitude~\cite{hennessy2019new,chien201110x10}.
Despite the well-known potential of hardware accelerators, their use in practice tends to be limited to a few application domains (ML~\cite{chen2016eyeriss, jouppi2017datacenter}, cryptography~\cite{samardzic2021f1}, audio/video codecs) where computational kernels can be easily re-used across a wide range of workloads.
The high levels of manual effort involved in designing hardware accelerators have been a significant impediment to their broader use.
Advances in abstraction and automation such as high-level synthesis (HLS)~\cite{martin2009high, lahti2018we} have addressed this challenge in part by enabling software-like specifications to be converted into register-transfer level (RTL) implementations. Nevertheless, designing a hardware accelerator for complex application workloads still requires significant effort as the designer needs to perform various steps such as profiling the application workload to identify target kernels for acceleration, re-writing the code to separate out the identified kernels, ensuring that the kernels adhere to the subset of programming constructs and data types supported by the HLS tool, and running the HLS tool to generate the RTL implementation. Notably, these steps require designers with considerable expertise in hardware design, micro-architecture and HLS tool use, making them challenging for application domain experts and software developers.
Recent advances in AI have led to AI agents that can autonomously set goals, plan and execute them with minimal or no human intervention~\cite{yao2022react}. These advances have enabled unprecedented levels of automation in a wide range of tasks, from generating natural language\cite{zhao2023survey}, images\cite{bie2024renaissance} and video\cite{team2025kling,gao2025seedance} to code generation\cite{jiang2026survey} and even scientific research\cite{novikov2025alphaevolvecodingagentscientific}. We explore the use of agentic AI to realize end-to-end automation of all the steps involved in hardware accelerator design.
Realizing this vision presents significant challenges. Despite recent advances in agentic and reasoning capabilities of AI systems, the underlying large language models (LLMs) are still prone to a wide range of inadequacies such as hallucination~\cite{huang2025survey}, laziness and context rot\cite{bai2024longbench}. Furthermore, since LLMs are trained on public data from the Internet, they are typically not exposed to the capabilities and usage of commercial EDA tools. Given the complexity of the task at hand, architecting the agentic flow in a manner that is cognizant of the limitations of artificial intelligence systems and the potential pitfalls they could face is vital to achieve reliable end-to-end automation of the accelerator design flow.
To that end, we present A3D, an agentic AI flow for end-to-end automated accelerator design. Starting with the codebase for an application (C/C++/CUDA), A3D executes and analyzes the workload, determines kernels that are performance bottlenecks, re-factors the code to extract the identified bottleneck kernels, prepares the kernels to be HLS-ready, and runs the HLS tool to generate the micro-architectural design (register-transfer level) for diverse accelerator designs spanning the speed-area tradeoff space.
To address the aforementioned challenges, the agentic architecture of A3D has the following key features:
\begin{itemize}
    \item A judicious partitioning of the tasks involved in end-to-end accelerator design (workload analysis, profiling, refactoring for HLS compatibility, etc.) into a set of steps with a dedicated (specialist) agent assigned to each step. 
    \item The agents employ pre-existing tools as well as custom tools that we implemented using the model context protocol (MCP) to strike a balance between using generative capabilities of the underlying LLM and the reliable functionality of deterministic tools. This careful delineation boosts the capabilities of the agents beyond what an LLM could achieve.
    \item Furthermore, each specialist agent is overseen by a dedicated verifier agent that iterates until the specialist agent completes the tasks correctly.
    \item A3D uses agentic RAG with a two-stage retrieval pipeline using state-of-the-art embedding and reranking models to explore the codebase and proprietary EDA tool documentation. 
\end{itemize}

\section{Related Works}
\label{sec:related}
\begin{table*}[!t]
\centering
\begin{tabular}{lcccccccc}
\toprule
\textbf{Framework} &
\textbf{\begin{tabular}[c]{@{}c@{}}Application\\Analysis\end{tabular}} &
\textbf{\begin{tabular}[c]{@{}c@{}}Workload\\Profiling\end{tabular}} &
\textbf{\begin{tabular}[c]{@{}c@{}}Kernel\\Extraction\end{tabular}} &
\textbf{\begin{tabular}[c]{@{}c@{}}Code\\Refactoring\end{tabular}} &
\textbf{\begin{tabular}[c]{@{}c@{}}Numerics\\Optimization\end{tabular}} &
\textbf{\begin{tabular}[c]{@{}c@{}}HLS\\Synthesis\end{tabular}} &
\textbf{\begin{tabular}[c]{@{}c@{}}Design Space\\Exploration\end{tabular}} &
\textbf{\begin{tabular}[c]{@{}c@{}}GPU\\Workloads\end{tabular}}\\
\midrule
C2HLSC~\cite{c2hlsc} & -- & -- & -- & \checkmark & -- & \checkmark & -- & -- \\
SAGE-HLS~\cite{sage-hls}        & -- & -- & -- & \checkmark & -- & -- & -- & -- \\
ChatHLS~\cite{chathls}          & -- & -- & -- & -- & -- & \checkmark & \checkmark & -- \\
LIFT~\cite{lift}                & -- & -- & -- & -- & -- & -- & \checkmark & -- \\
LLM-DSE~\cite{llm-dse}          & -- & -- & -- & -- & -- & -- & \checkmark & -- \\
HLSPilot~\cite{hlspilot}        & -- & -- & -- & \checkmark & -- & \checkmark & \checkmark & -- \\
HLS-Repair~\cite{hls-repair} & -- & -- & -- & \checkmark & \checkmark & \checkmark & -- & -- \\
HLSRewriter~\cite{hlsrewriter} & -- & -- & -- & \checkmark & -- & \checkmark & -- & -- \\
Agentic-HLS~\cite{oztas2024agentic} & -- & -- & -- & -- & -- & -- & \checkmark & -- \\
\midrule
\textbf{A3D (This work)}         & \checkmark & \checkmark & \checkmark & \checkmark & \checkmark & \checkmark & \checkmark & \checkmark \\
\bottomrule
\end{tabular}
\caption{Coverage of HLS workflow stages across related frameworks. Existing work addresses individual downstream stages; A3D is the first to automate the full pipeline from application analysis through design space exploration.}
\label{tab:related-work-coverage}
\end{table*}

Recent works have developed specialized LLM-driven frameworks targeting individual stages of the HLS-based accelerator design workflow. We organize them by the stage they address and then identify the gap that A3D fills.

\paragraph{Code refactoring for synthesizability.}
Translating standard C/C++ into the synthesizable subset accepted by HLS tools is a labor-intensive prerequisite that several frameworks now automate.
SAGE-HLS~\cite{sage-hls} fine-tunes a specialized model on a novel Verilog-to-C dataset using an AST-guided training objective, enabling the model to learn synthesis-friendly code structure rather than surface-level text patterns.
C2HLSC~\cite{c2hlsc} takes a tool-in-the-loop approach: it iteratively submits refactored code to the HLS compiler, parses error logs, and re-prompts the LLM until synthesis succeeds.
HLS-Repair~\cite{hls-repair} similarly closes the loop with the HLS tool but additionally addresses numeric type optimization, converting floating-point operations to fixed-point representations to improve hardware efficiency.
HLSRewriter~\cite{hlsrewriter} extends C2HLSC's iterative approach with a multi-pass refactoring pipeline that handles a broader set of unsupported constructs, including pointer arithmetic and dynamic memory, achieving higher synthesis success rates. HLSDebugger~\cite{wang2025hlsdebugger} targets the downstream problem of identifying and correcting logic bugs in HLS code using LLM-based solutions.

\paragraph{Design space exploration.}
Once a kernel is synthesizable, selecting the right combination of optimization directives (loop unrolling factors, pipeline initiation intervals, array partitioning schemes) remains an expert-driven art due to the large design space~\cite{schafer2019high}.
LIFT~\cite{lift} tackles this with a hybrid LLM--GNN architecture: a Graph Neural Network reasons over control and data dependencies to supervise the LLM's fine-tuning for pragma insertion, substantially outperforming both heuristic-based DSE tools and na\"ive general-purpose LLM insertions.
LLM-DSE~\cite{llm-dse} deploys a multi-agent team that collaboratively navigates the design space; the Critic agent parses HLS synthesis reports and provides natural-language feedback to guide subsequent iterations.
ChatHLS~\cite{chathls} separates concerns into an HLSFixer agent for automated error correction and an HLSTuner agent that iteratively refines pragma configurations through synthesis-report feedback.
Agentic-HLS~\cite{oztas2024agentic} combines an LLM with a GNN in a chain-of-thought reasoning framework to predict HLS quality-of-result metrics before synthesis, enabling rapid design space pruning.

\paragraph{End-to-end orchestration.}
HLSPilot~\cite{hlspilot} is the closest prior art to A3D, orchestrating profiling, C-to-HLS translation, DSE, and host-side code generation in a four-stage pipeline augmented with retrieval-augmented generation from Xilinx Vitis documentation.
However, HLSPilot, and all frameworks above, share a critical assumption: they begin with isolated, well-defined kernel functions where the performance bottleneck has already been extracted into a standalone test harness and are not designed to handle the complexity of full application analysis, kernel identification and extraction.

The tasks that \emph{precede} this starting point remain unaddressed in all existing frameworks: analyzing HPC-scale applications with hundreds of source files, profiling representative workloads, identifying bottleneck kernels buried within deep class hierarchies, resolving their data-structure dependencies across multiple abstraction layers, and isolating them into self-contained harnesses. Table~\ref{tab:related-work-coverage} summarizes this coverage gap. A3D closes this gap by automating these aforementioned upstream tasks along with the downstream tasks like refactoring, HLS synthesis and design space exploration, thus achieving end-to-end automation of going from application to accelerator. Additionally, most scientific applications are accelerated on GPUs despite being poorly served by the programmable parallelism offered by GPUs and could benefit from custom accelerators. None of the existing frameworks address such scenarios, where performance bottlenecks reside in CUDA kernels. A3D is capable of handling CUDA bottlenecks and developing custom accelerator designs directly from CUDA code.

\section{Agentic AI flow for Accelerator Design}
\label{sec: A3D}

In this section, we present A3D, an Agentic AI flow for Automated Accelerator Design. We present the technical details and design choices behind A3D, and how it leverages the agentic capabilities of LLMs while addressing their limitations.

The A3D flow starts with the application codebase and a workload to execute on the application, along with general natural language instructions regarding compilation, execution and environment setup. An {\it application} is typically a software program that was designed to run on general-purpose computing platforms such as CPUs and GPUs. The application might contain thousands of source files typically written in a high-level programming language such as C/C++, CUDA or Python to meet the functional and performance requirements of the target domain. In this work, we focus on scientific computing applications written in C/C++ for molecular dynamics simulations and quantum chemistry. A {\it workload} is an input configuration that exercises the application to perform a specific task or calculation. A workload typically contains input data files and requirements for the application to run. The workloads can vary significantly in terms of data size, complexity, and computational requirements and the workloads studied in this work are typically run on high-performance computing (HPC) systems.

\begin{figure*}[b]
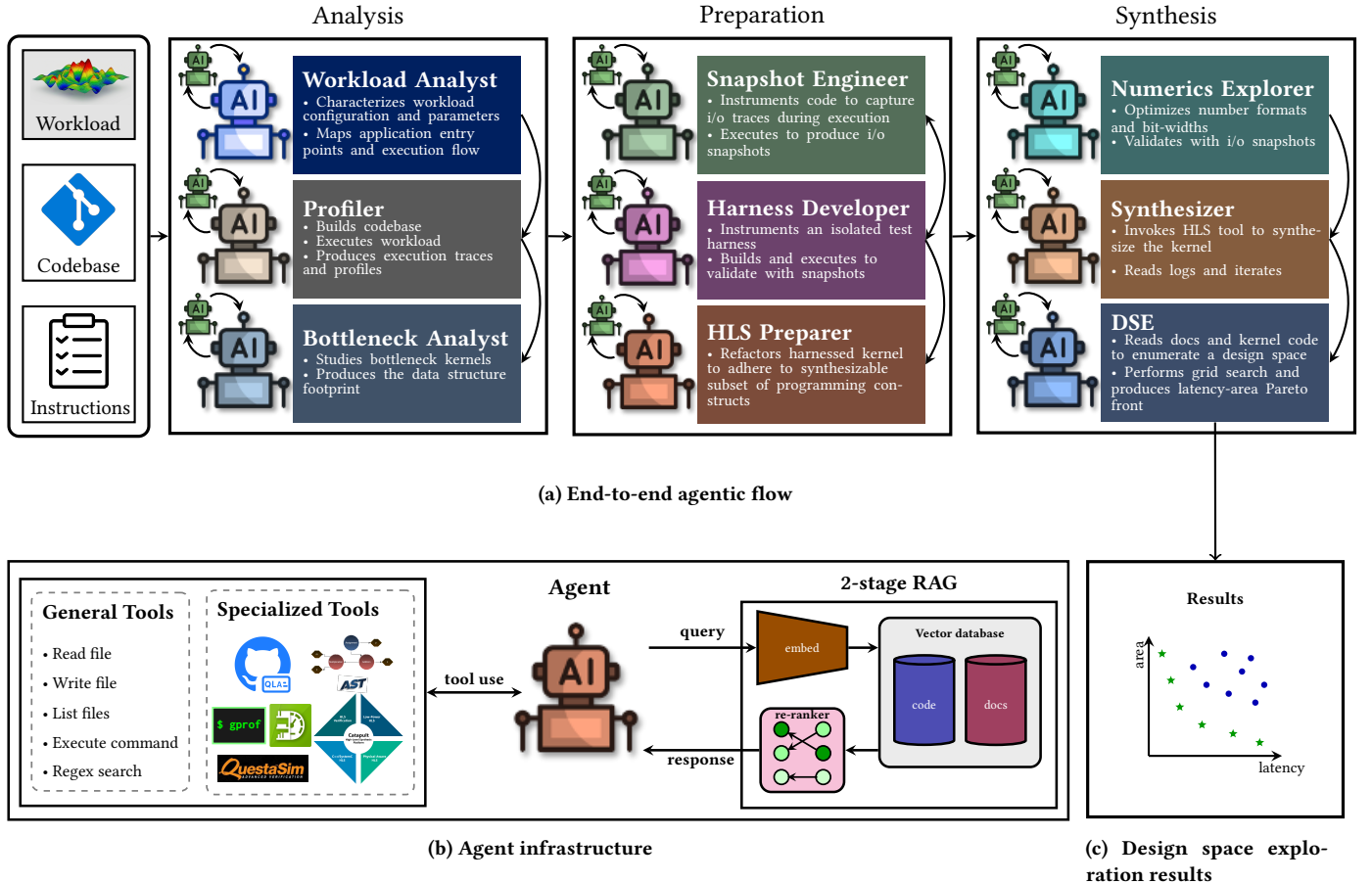

    \centering
    \begin{subfigure}{\textwidth}
        \centering
        \input{sections/figs/A3D.tex}
        \caption{End-to-end agentic flow}
    \end{subfigure}
    
    \vspace{1em}
    
    \begin{subfigure}[t]{0.81\textwidth}
        \centering
        \resizebox{\linewidth}{!}{\input{sections/figs/infrastructure.tex}}
        \caption{Agent infrastructure}
    \end{subfigure}
    \hfill
    \begin{subfigure}[t]{0.181\textwidth}
        \centering
        \begin{tikzpicture}[
    remember picture,
    scale=0.6,
    transform shape,
    baseline=(current bounding box.north),
    >=stealth,
    results box/.style={
        draw=black,
        very thick,
        inner sep=20pt,
        rectangle,
        minimum width=5.75cm,
        minimum height=5.75cm
    }
]

\node[results box] (container) {};

\node[above=0cm of container.center, yshift=1.8cm, font=\bfseries\Large] {Results};

\draw[->, thick] ([xshift=-1.5cm, yshift=-1.5cm]container.center) -- ([xshift=1.5cm, yshift=-1.5cm]container.center) node[pos=1.0, below, font=\large] {latency};
\draw[->, thick] ([xshift=-1.5cm, yshift=-1.5cm]container.center) -- ([xshift=-1.5cm, yshift=1.2cm]container.center) node[pos=1.0, left, font=\large, rotate=90, yshift=0.2cm] {area};

\filldraw[fill=blue!70!black, draw=none] ([xshift=-0.5cm, yshift=0.5cm]container.center) circle (2pt);
\filldraw[fill=blue!70!black, draw=none] ([xshift=0.2cm, yshift=0.8cm]container.center) circle (2pt);
\filldraw[fill=blue!70!black, draw=none] ([xshift=0.6cm, yshift=0.4cm]container.center) circle (2pt);
\filldraw[fill=blue!70!black, draw=none] ([xshift=0.8cm, yshift=0.7cm]container.center) circle (2pt);
\filldraw[fill=blue!70!black, draw=none] ([xshift=1.1cm, yshift=0.1cm]container.center) circle (2pt);
\filldraw[fill=blue!70!black, draw=none] ([xshift=0.3cm, yshift=-0.1cm]container.center) circle (2pt);
\filldraw[fill=blue!70!black, draw=none] ([xshift=-0.2cm, yshift=0.1cm]container.center) circle (2pt);
\filldraw[fill=blue!70!black, draw=none] ([xshift=0.9cm, yshift=-0.3cm]container.center) circle (2pt);

\def\starpath{ (90:3pt) -- (162:1.1pt) -- (234:3pt) -- (306:1.1pt) -- (378:3pt) -- (450:1.1pt) -- (522:3pt) -- (594:1.1pt) -- (666:3pt) -- cycle }
\tikzset{
    green star/.style={
        fill=green!60!black,
        draw=none
    }
}

\begin{scope}[shift={([xshift=-1.2cm, yshift=0.8cm]container.center)}] \filldraw[green star] \starpath; \end{scope}
\begin{scope}[shift={([xshift=-1.0cm, yshift=0.2cm]container.center)}] \filldraw[green star] \starpath; \end{scope}
\begin{scope}[shift={([xshift=-0.8cm, yshift=-0.4cm]container.center)}] \filldraw[green star] \starpath; \end{scope}
\begin{scope}[shift={([xshift=-0.3cm, yshift=-0.8cm]container.center)}] \filldraw[green star] \starpath; \end{scope}
\begin{scope}[shift={([xshift=0.4cm, yshift=-1.0cm]container.center)}] \filldraw[green star] \starpath; \end{scope}
\begin{scope}[shift={([xshift=1.0cm, yshift=-1.2cm]container.center)}] \filldraw[green star] \starpath; \end{scope}


\end{tikzpicture}
        \caption{Design space exploration results}
    \end{subfigure}
    
    \tikz[remember picture, overlay] \draw[->, thick] (dse_text.south) to[out=-90, in=90] (container.north);
    
    \caption{A3D: Agentic AI flow for Accelerator Design. (a)~End-to-end multi-agent pipeline spanning three phases: Analysis, Preparation, and Synthesis (b)~Shared agent infrastructure including general-purpose and specialized tools alongside a two-stage RAG pipeline over the codebase and EDA tool documentation. (c)~Latency--area design space produced by the DSE agent providing a diverse set of Pareto-optimal accelerator designs.}
    \label{fig:A3D}
\end{figure*}

A3D decomposes the complex process of accelerator design into three major phases, as illustrated in Figure~\ref{fig:A3D}: (1) the Analysis phase, where the application and workload are analyzed to identify performance bottlenecks and the dependencies of the bottlenecks are determined, (2) the Preparation phase, where the identified bottlenecks are isolated from the application codebase and refactored to be compatible with HLS tools, and (3) the Synthesis phase, where the prepared bottlenecks are synthesized using an HLS tool to generate hardware accelerator designs. Each phase is further decomposed into smaller manageable tasks that are assigned to specialist agents designed to perform these tasks effectively. The cross-cutting infrastructure principles of tool augmentation, agentic RAG, adversarial verification, and orchestration are described in \S\ref{sec:infrastructure}.

\subsection{Analysis Phase}
\label{sec:analysis}
The algorithmic and computational complexity of scientific computing applications often lead to performance bottlenecks that limit their overall performance on programmable general-purpose computing platforms. These bottlenecks can arise from various factors such as complexity of algorithms, suboptimal data structures, memory access patterns, and unsuitability for the traditional SIMD or SIMT parallelism paradigms employed by CPUs and GPUs. Identifying these bottlenecks and building accelerators to offload them can significantly improve the overall performance of the application. However, identifying these bottlenecks in large complex applications is a challenging task that requires expertise in both the application domain and software development, as the bottlenecks can be deeply embedded within the application codebase with complex dependencies. 

A3D addresses this challenge with three specialist agents that work in sequence to identify the bottlenecks effectively. The first agent, the {\it Workload Analyst}, analyzes the application codebase using semantic search and the workload to develop an understanding of the application's functionality and the workload's characteristics. The second agent, the {\it Performance Profiler}, uses a profiling tool like gprof to profile the application with the given workload and identify the functions that consume the most execution time. The third agent, the {\it Bottleneck Analyst}, builds on the insights from the previous two agents to analyze the identified performance-critical functions and identify their dependencies within the application codebase. Determining the list of data structure dependencies is critical for accurate implementation of the bottleneck kernels. In order to ensure that the Bottleneck Analyst can reliably and repeatably produce the exhaustive list of dependencies, a static analysis tool is implemented using CodeQL. The Bottleneck Analyst uses said \texttt{codeQLAnalysis} tool to perform static analysis of the codebase and identify the data structure and function dependencies of the performance-critical functions. The Bottleneck Analyst then produces a detailed report that summarizes the identified bottlenecks and provide valuable insights into the data types, sizes, and structures used by these performance-critical functions.

\subsection{Preparation Phase}
\label{sec:preparation}
High level synthesis is a widely used methodology for designing hardware accelerators from high-level programming languages such as C/C++. HLS tools can automatically generate hardware description language (HDL) code from high-level C/C++ code, significantly reducing the development time and effort required for hardware design. However, HLS tools typically support only a subset of the C/C++ language features and constructs, and certain programming patterns may not be synthesizable. The performance bottlenecks identified in the previous step can be expected to be highly optimized for the target general-purpose computing platform, and thus, may use complex programming constructs and patterns that are not compatible with HLS tools like multi-dimensional pointers, dynamic memory, STL containers etc. Therefore, the identified bottlenecks must be refactored to ensure compatibility with HLS tools while preserving their functionality. 

A3D addresses this challenge with a three step process. The first step involves the {\it Snapshot Engineer} agent that utilizes the data structure dependencies identified by the Bottleneck Analyst to take snapshots of the input and output datastructures of the bottleneck functions during workload execution. The second step involves the {\it Harness Developer} agent that implements an isolated test harness for the bottleneck function and validates its correctness using the input and output snapshots generated in the previous step. The test harness provides a controlled environment for testing and validating the bottleneck function, ensuring that it behaves as expected when refactored and optimized for HLS. The snapshot engineer and harness developer agents are designed to work in collaboration, with the snapshot engineer providing the necessary data structure coverage so that the harness developer can implement an effective test harness. If any discrepancies are found during validation, the agents iteratively refine data capture and test harness implementation until numerical correctness is achieved.

\begin{figure}[htbp]
\centering
\input{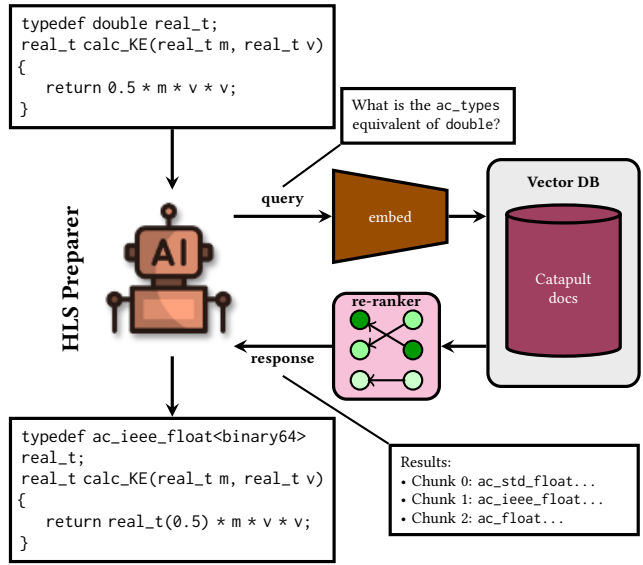}
\caption{HLS Preparer agent refactoring floating-point types to HLS-compatible types via agentic RAG. The agent queries the Catapult documentation vector database for the \texttt{ac\_types} equivalent of \texttt{double} and rewrites the source code with the retrieved replacement.}
\label{fig:hls_preparer_example}
\end{figure}

The third step involves the {\it HLS Preparer} agent that refactors the bottleneck function to ensure compatibility with HLS tools, by replacing unsupported programming constructs. Moreover, HLS tools provide their own libraries for optimized implementations of floating and fixed-point number formats, mathematical functions, and data structures. The HLS Preparer agent uses its agentic RAG capabilities (\S\ref{sec:infrastructure}) to understand the HLS-optimized counterparts of the number formats and math functions used in the bottleneck function. These necessary refactorings are quite complex and require deep understanding of both the algorithmic needs of the application and HLS tool capabilities. Therefore, A3D breaks down the HLS preparation task into smaller manageable sub-tasks, each handled by dedicated specialist agents. The HLS Preparer agent orchestrates this process by collaborating with the specialist agents. For example, Figure~\ref{fig:hls_preparer_example} illustrates how an HLS Preparer agent refactors standard floating point data types to HLS-compatible data types by performing a natural language query to identify the appropriate replacements. To further streamline this process, AST-based tools (\S\ref{sec:infrastructure}) automate common refactorings such as pointer to array conversions and literal typecastings. Specialist agents leverage these tools to perform the required refactorings in a systematic manner.

At the end of the preparation phase, A3D produces an HLS-compatible version of the bottleneck function, with statically defined data structures and HLS-compatible number formats and math functions, along with a validated test harness for the function.

\subsection{Synthesis Phase}
\label{sec:synthesis}
One of the key advantages of building custom hardware accelerators is the ability to choose number formats and bitwidths that are tailored to the application's requirements and to build the necessary data paths and compute units accordingly. This flexibility allows for significant improvements in performance, area, and energy efficiency compared to general-purpose computing platforms that use fixed number formats and bitwidths. Thus, as part of the HLS phase, the {\it Numerics Explorer} agent explores different number formats and bitwidths for each data element in the bottleneck function with the goal of minimizing bitwidths while preserving the functional correctness of the application. While the numerics explorer takes a trial-and-error approach in this work by iteratively reducing the bitwidths, it should be feasible to create tools based on formal methods that the agent can leverage to perform this task more systematically.

Once the number formats and bitwidths have been determined, the {\it Synthesizer} agent performs high-level synthesis of the prepared bottleneck function using the Catapult HLS tool. In case of synthesis failures, the synthesizer agent analyzes the synthesis reports and error logs, identifies the root causes of the failures, and finds appropriate solutions using its agentic RAG capabilities to explore the HLS tool documentation. The synthesizer agent iteratively refines the HLS-compatible code to achieve successful synthesis.

HLS tools provide several optimization directives, such as loop unrolling, pipelining, and array partitioning, that can be used to explore the speed-area tradeoff space of potential hardware accelerator designs. The {\it Design Space Enumerator} agent studies the bottleneck code and the optimization directives supported by the HLS tool to enumerate a diverse set of optimization directives and their configurations that can be applied to the bottleneck function during synthesis.

The DSE agent encodes its analysis into a structured YAML specification that declaratively describes the optimization dimensions (e.g., design goal, clock period, loop unrolling factors) and their option values, along with any implied constraints such as memory interleaving requirements. A3D provides DSE generation tools that parse this YAML specification, perform combinatorial expansion across all dimensions, and emit one HLS directive (Tcl) file per design point along with a summary manifest. This deterministic tool chain ensures that every generated configuration is syntactically correct, a task that would be highly error-prone if performed generatively by the LLM across hundreds of configurations.

A grid search is then performed by synthesizing every design point, reporting the performance and area metrics for each, and extracting the Pareto-optimal set of designs. A3D provides a DSE execution tool that manages the parallel synthesis of all design points, handling job scheduling across local and remote compute nodes, Catapult invocation with the appropriate directive file for each design point, and log collection. The grid search is embarrassingly parallel and provides complete coverage of the feasibility boundary between viable and infeasible designs. A3D currently employs this grid search strategy for design space exploration, but more advanced techniques such as Bayesian optimization~\cite{sohrabizadeh2022autodse} or evolutionary algorithms~\cite{schafer2019high} can be integrated into the DSE execution tool to improve the efficiency of design space exploration.

\subsection{Agent Infrastructure}
\label{sec:infrastructure}

The specialist agents described above are supported by a shared infrastructure layer comprising custom tools, an agentic RAG pipeline, adversarial verification loops, and a top-level orchestrator. This infrastructure is critical to achieving reliable end-to-end automation despite the inherent limitations of LLMs.

\subsubsection{Custom and Pre-existing Tools}
The agents leverage general-purpose tools for file system exploration, reading and writing to files, and running shell commands. A3D also invokes pre-existing tools such as gprof for profiling and the Catapult HLS tool for high-level synthesis. Additionally, A3D implements task-specific custom tools (Table~\ref{tab:custom-tools}) for code analysis using CodeQL, code refactoring using Abstract Syntax Tree (AST) transformations and helper tools for performing design space exploration. These tools provide deterministic, repeatable transformations that complement the generative capabilities of the LLM agents, eliminating a significant source of errors in mechanical code transformations.

\begin{table}[t]
\centering
\caption{Custom tools implemented for A3D. These tools provide deterministic, repeatable transformations that complement the generative capabilities of the LLM agents.}
\label{tab:custom-tools}
\small
\begin{tabular}{p{2cm}p{1.2cm}p{4.5cm}}
    \toprule
    \textbf{Tool} & \textbf{Phase} & \textbf{Function} \\
    \midrule
    \texttt{codeQL\-Analysis} & Analysis & Static analysis of codebase to extract data-structure and function dependencies of bottleneck kernels \\
    \texttt{staticMem} & Preparation & AST-based conversion of dynamic memory (multidimensional pointers, \texttt{new[]}) to statically-sized C-style arrays with compile-time constants \\
    \texttt{literal\-Typecast} & Preparation & AST-based insertion of explicit type casts on all numeric literals to prevent implicit conversion ambiguities \\
    \texttt{genDSE} & Synthesis & Parses YAML design space specification, performs combinatorial expansion across all dimensions, emits per-design-point Tcl directive files \\
    \texttt{runDSE} & Synthesis & Parallel synthesis execution across local and remote compute nodes with job scheduling, failover, and log collection \\
    \bottomrule
\end{tabular}
\end{table}

\subsubsection{Agentic RAG Pipeline}
A3D employs a two-stage retrieval pipeline to enable agentic RAG capabilities for the agents to semantically explore the application codebase and the proprietary EDA tool documentation. For example, when an agent encounters an unfamiliar construct or needs to identify the HLS-compatible counterpart of a standard library function, it issues a natural language query that is converted into an embedding, matched against vector databases populated with the codebase and tool documentation, and re-ranked for relevance. The top results are provided to the agent as context for the assigned task. This pipeline is essential for bridging the gap between the LLM's training data and the proprietary knowledge required for hardware design.

\subsubsection{Specialist--Verifier Loop}

\begin{figure}
    \centering
    \input{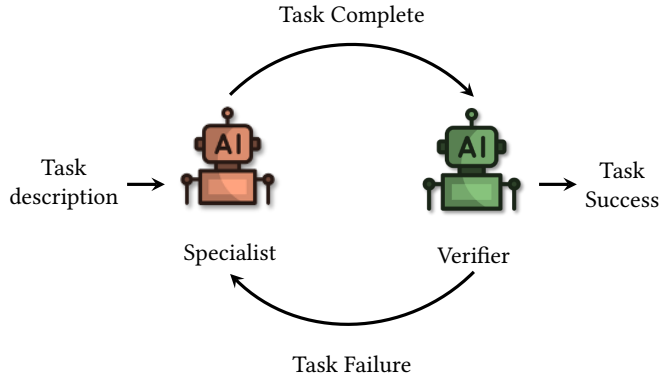}
    \caption{Specialist--Verifier iterative process loop. The verifier adversarially evaluates the specialist's deliverables and returns critical feedback on errors or shortcomings, looping until quality standards are met.}
    \label{fig:specialist_verifier_loop}
\end{figure}

Given the complexity of the tasks involved in accelerator design and the stochastic nature of LLMs, it is inevitable that the outputs produced by the specialist agents may contain errors or may not meet the desired quality standards or may not meet the expected list of deliverables. To add robustness to the A3D flow and eliminate the need for human intervention, A3D employs an iterative process loop with specialist and verifier agents. Verifier agents are designed to adversarially evaluate the outputs produced by specialist agents, identify any errors, shortcomings or false claims, and provide critical feedback to the specialist agents for task resumption and refinement. Figure~\ref{fig:specialist_verifier_loop} illustrates this process loop. The transitions between the specialist and verifier agents are automatic and the loop continues until the verifier agent validates that the deliverables produced by the specialist agent meet the desired outcomes. While this process loop can potentially lead to longer task completion times due to multiple iterations, we empirically find that it significantly improves the quality of the outputs and significantly reduces the need for human intervention.

\subsubsection{Orchestration}
To orchestrate this end-to-end flow with full autonomy, A3D employs the {\it Interfacer} agent, the top-level orchestrator. The Interfacer sequences the three phases, passes accumulated context (analysis reports, extracted artifacts, validation results) from one phase to the next, and re-invokes specialist agents as required. Crucially, the Interfacer operates at the phase level; within each phase, individual agents may themselves coordinate sub-tasks (as described above for HLS preparation), but the Interfacer is responsible for the overall flow progression and inter-phase handoff.

\section{Results}
\label{sec:results}

In this section, we present the results of evaluating A3D. First, we characterize the benchmarks and the complexity of the kernels identified by A3D from real-world scientific applications (\S\ref{sec:kernel-complexity}). We then present the end-to-end pipeline results (\S\ref{sec:pipeline-results}) demonstrating that A3D autonomously transforms complex scientific codebases into synthesized hardware accelerator designs, and compare A3D-generated accelerator designs against an edge GPU (\S\ref{sec:gpu-comparison}). Next, we present a rigorous ablation study that quantifies the contribution of each architectural layer in A3D's HLS preparation phase (\S\ref{sec:ablation}), and finally walk through the end-to-end agentic flow qualitatively (\S\ref{sec:e2e-flow}).

\subsection{Benchmarks and Kernel Characteristics}
\label{sec:kernel-complexity}

We evaluate A3D on two flagship scientific computing applications: LAMMPS~\cite{LAMMPS}, a molecular dynamics (MD) simulation toolkit, and QMCPACK~\cite{qmcpack}, a quantum Monte Carlo (QMC) application. LAMMPS simulates systems of atoms that interact with each other through force functions called \textit{interatomic potentials}: at each timestep the simulator evaluates forces on every atom and updates their positions. The force evaluation is the most computationally expensive step and the primary target for hardware acceleration. QMCPACK simulates the behavior of electrons in materials by stochastically sampling a mathematical function (the \textit{wavefunction}) that describes the quantum state of the system; the computational cost is dominated by evaluating particle--particle interaction energies across the simulated system. Both applications are large-scale C++ codebases with hundreds of source files, deep class hierarchies, and pervasive use of dynamic memory, templates, and pointer-based data structures. These are representative of the software complexity that A3D is designed to handle. We use representative workloads for both applications, scaled down to run on a single CPU core for the purpose of demonstrating the full workflow.

\subsubsection{LAMMPS Kernels}
Five LAMMPS workloads were selected to span a progression of interatomic potential complexity: \textit{lj}, \textit{eam}, \textit{rhodo}, \textit{snap}, and \textit{reaxff}. The simplest (\textit{lj}) computes forces based only on the distance between pairs of atoms; the most complex (\textit{reaxff}) models the formation and breaking of chemical bonds, requiring interactions that simultaneously involve four atoms. The performance-critical kernels identified by A3D span two fundamental tasks: force evaluation and neighbor-list construction.

\texttt{PairLJCut::Compute()} evaluates the Lennard-Jones pair potential, the simplest widely-used force function, where the force between two atoms depends only on their distance. A cutoff distance is imposed so that only nearby pairs interact, which avoids the $O(N^2)$ cost of evaluating all pairs but requires a \textit{neighbor list} data structure that records, for each atom, which other atoms are close enough to interact. The kernel iterates over each atom's neighbor list, an irregular, input-dependent data structure, computing pairwise distances, forces, and energies. A physics symmetry (equal and opposite forces between each pair) halves the computation but introduces concurrent read-modify-write updates to force arrays, and the distance-based cutoff leads to irregular control flow. This kernel dominates the execution time in the \textit{lj} workload.

\texttt{NPairBin::Build()} constructs the neighbor list itself, a spatial indexing structure shared by all force-evaluation kernels. The algorithm partitions the simulation domain into spatial bins and, for each atom, tests candidates from adjacent bins against a distance threshold. This kernel is characterized by multi-level pointer indirection (\texttt{int**} neighbor arrays with runtime-determined dimensions), dynamic output sizes (each atom has a different number of neighbors depending on local conditions), and indirect memory addressing through bin-to-atom mappings. It is the dominant bottleneck in the \textit{eam} workload (which simulates metallic systems where force computation is less expensive relative to neighbor-list construction) and the \textit{rhodo} workload (which simulates a biological protein).

\texttt{SNA::compute\_uarray()} computes mathematical descriptors of each atom's local neighborhood for a machine-learned force function (SNAP). Instead of using a fixed formula, SNAP uses descriptors that are projections of the neighbor arrangement onto a set of basis functions as inputs to a learned model that predicts forces. The descriptor computation involves deeply nested loops over basis function indices, with complex-valued arithmetic and transcendental functions (\texttt{sin}, \texttt{cos}) in the inner loop. A configuration parameter (\texttt{twojmax}$= 8$) directly controls the loop depth and, consequently, the computational intensity. While compact in source lines, the mathematical density and loop nesting make this kernel a challenging target for HLS.

\texttt{ReaxFF::Torsion\_Angles()} evaluates the energy contribution from groups of four bonded atoms in a \textit{reactive} force field that can model the formation and breaking of chemical bonds. This capability requires tracking the bonding state of every atom and evaluating multi-body interaction terms (two-body, three-body, and four-body). The torsion angle term computes the energy penalty for deviations from preferred dihedral angles across four bonded atoms ($i$-$j$-$k$-$l$). Because bonds can form and break during the simulation, the set of valid four-atom groups is determined dynamically from the evolving bond topology, producing a four-deep nested loop hierarchy: the outermost loop iterates over all 2,048 atoms, the second over each atom's bond partners ($\sim$20 per atom), the third over angle partners ($\sim$10), and the innermost over torsion partners ($\sim$10). The innermost loop body contains $\sim$200 floating-point operations including transcendental functions. The kernel exhibits K=4 windowed memory access patterns on the atom position array (simultaneously accessing atoms $i$, $j$, $k$, and $l$), accumulator dependencies on force and energy arrays, and indirect indexing through bond and interaction lists with runtime-determined bounds. This combination of deep nesting, irregular memory access, and accumulation patterns makes it among the most challenging kernels for HLS synthesis considered to-date.

\subsubsection{QMCPACK Kernels}
QMCPACK uses a stochastic algorithm to compute the energy of electrons in a material. A key computational challenge is evaluating the electrostatic interaction between charged particles: because this interaction extends to infinite range (unlike the cutoff-based forces in LAMMPS), it cannot be computed by simply summing over nearby pairs. Instead, the interaction is calculated through Ewald summation, which splits it into a short-range part (handled in real space) and a long-range part (handled in frequency space via Fourier transforms). The two kernels identified by A3D both arise from the frequency-space (long-range) part of this computation. Rather than targeting the dense linear algebra (BLAS) operations that dominate many quantum chemistry codes, we prioritize these non-BLAS-like kernels with irregular computation patterns to better demonstrate the breadth of A3D's capabilities.

\texttt{LRHandlerBase::Evaluate()} computes the long-range energy contribution. The Fourier-space coefficients are pre-computed; the kernel itself is a pure accumulation loop that computes weighted dot products of pre-computed arrays grouped into shells of varying size. For each shell, the inner loop accumulates element-wise products across a variable number of elements, and the outer loop multiplies each shell's sum by a pre-computed weight.

\texttt{StructFact::computeRhok()} computes the frequency-space representation of the particle distribution, which serves as the input to the energy evaluation above. For each particle, the kernel computes the dot product of the particle's position with each frequency-space vector, evaluates \texttt{cosine} - \texttt{sine} pairs of the result, and accumulates the complex exponential into per-species output arrays. The kernel processes frequency-space vectors in blocks of 512, with the block count and final block size determined at runtime. The HLS challenges include CORDIC-based transcendental functions, species-based indirect write targeting (a particle's species ID determines which output array receives the accumulation), and data-dependent loop bounds with early-exit semantics for the blocked processing.

\subsubsection{Comparison with Prior Work}
Beyond the workflow coverage gap discussed in \S\ref{sec:related} (Table~\ref{tab:related-work-coverage}), the kernels themselves present qualitatively different challenges from the benchmarks used by prior LLM-driven HLS frameworks. C2HLSC~\cite{c2hlsc}, SAGE-HLS~\cite{sage-hls}, HLSPilot~\cite{hlspilot}, LIFT~\cite{lift}, and LLM-DSE~\cite{llm-dse} are evaluated on isolated benchmark kernels drawn from suites such as Polybench, MachSuite~\cite{reagen2014machsuite}, and CHStone~\cite{hara2008chstone}. These benchmarks typically consist of compact loop nests with regular, statically determinable access patterns, fixed array dimensions, and no external data structure dependencies.

In contrast, the kernels processed by A3D exhibit several dimensions of complexity absent from these benchmark suites. \emph{Data-structure depth:} functions operate on class member variables accessed through multi-level pointer indirection (up to \texttt{****fbp} in ReaxFF), arrays of nested structs containing struct members (e.g., \texttt{bond\_data} embedding a \texttt{bond\_order\_data} struct with 20 fields), and complex data dependencies. \emph{Runtime-determined sizing:} array dimensions are not compile-time constants but are derived from simulation parameters, requiring snapshot-based size discovery and static over-provisioning (e.g., 168 atoms $\to$ 2,048-element arrays). \emph{Incompatible programming patterns:} the kernels make heavy use of dynamic memory allocation, pointer arithmetic, STL containers, and virtual dispatch. These patterns are fundamentally incompatible with HLS tools and require systematic refactoring across dozens of constructs. \emph{GPU-native workloads:} one kernel (\texttt{k\_lj\_fast}) was extracted from a CUDA implementation, requiring translation of GPU-specific idioms (thread indexing, shared memory, warp-level operations) into HLS-compatible sequential C++.

\subsection{End-to-End Pipeline Results}
\label{sec:pipeline-results}

A3D successfully processed all seven kernels identified in \S\ref{sec:kernel-complexity} through the complete pipeline with no human intervention. Table~\ref{tab:pipeline-summary} presents a comprehensive summary of the end-to-end pipeline results for each kernel, spanning all three phases of the A3D flow: the analysis phase (bottleneck identification and dependency extraction), the preparation phase (snapshot capture, harness development, and HLS-compatible refactoring), and the synthesis phase (numerics optimization, HLS synthesis, and design space exploration).

\begin{table*}[t]
\centering
\caption{End-to-end pipeline results for all kernels processed by A3D. Each kernel was autonomously taken through the full A3D flow, spanning workload analysis, bottleneck identification, code refactoring, HLS synthesis, and design space exploration, with zero human intervention.}
\label{tab:pipeline-summary}
\small
\begin{tabular}{llcccc}
    \toprule
    \textbf{Kernel} &
    \textbf{App} &
    \textbf{\begin{tabular}[c]{@{}c@{}}DSE\\points\end{tabular}} &
    \textbf{\begin{tabular}[c]{@{}c@{}}Successful\\syntheses\end{tabular}} &
    \textbf{\begin{tabular}[c]{@{}c@{}}Synthesis\\Success Rate\end{tabular}} &
    \textbf{\begin{tabular}[c]{@{}c@{}}Pareto-optimal\\designs\end{tabular}} \\
    \midrule
    \texttt{PairLJCut::Compute()}      & LAMMPS  & 1080 & 61  & 5.6\%  & 8  \\
    \texttt{NPairBin::Build()}         & LAMMPS  & 240 & 136 & 56.7\% & 15 \\
    \texttt{SNA::compute\_uarray()}    & LAMMPS  & 180 & 61  & 33.9\% & 11 \\
    \texttt{ReaxFF::Torsion\_Angles()} & LAMMPS  & 648 & 109 & 16.8\% & 7  \\
    \texttt{PairEAM::Compute()}        & LAMMPS  & 504 & 72  & 14.3\% & 7  \\
    \texttt{PairLJCut::k\_lj\_fast()}  & LAMMPS  & 486 & 295 & 60.7\% & 16 \\
    \texttt{StructFact::computeRhok()} & QMCPACK & 384 & 289 & 75.3\% & 15 \\
    \texttt{LRHandlerBase::Evaluate()} & QMCPACK & 54  & 49  & 90.7\% & 13 \\
    \bottomrule
\end{tabular}
\end{table*}

\subsubsection{Implementation}
\label{sec:implementation}

The agents in A3D are implemented using the Roo Code agent development toolkit~\cite{roocode}, which provides a framework for building and executing agentic workflows. While Roo Code supports choosing the underlying LLM for each agent, we use Claude Sonnet 4.5 as the underlying LLM for all agents in A3D due to its strong reasoning and code generation capabilities. The agentic RAG pipeline is implemented using state-of-the-art embedding and re-ranking models~\cite{zhang2025qwen3}, and vector databases. When the agent makes an enquiry, it is converted into a 4096-dimensional embedding using Qwen3-Embedding-8B. This embedding is then used to retrieve the top-k cosine-similar chunks from vector databases that have been populated with the codebase and the EDA tool documentation, respectively. The retrieved chunks are then re-ranked using Qwen3-Reranker-8B to reorder them based on their relevance to the original query. The top-m re-ranked chunks are then provided to the agent as context for the assigned task.
The HLS tool used in A3D is the Catapult HLS tool from Siemens, a widely used commercial HLS tool that supports C++ input and generates RTL implementations.

\subsubsection{Design Space Exploration}

A3D automatically identifies optimization opportunities for each kernel, such as loop unrolling, loop pipelining, and array partitioning factors, alongside global synthesis settings like target clock period and optimization goals (speed vs.\ area), and enumerates a combinatorial design space from these dimensions. Each design point is synthesized independently and a Pareto-optimal set of designs is extracted. Across all eight kernels, A3D enumerated a total of 3,216 design points and successfully synthesized 1,036 of them, yielding 92 Pareto-optimal designs (Table~\ref{tab:pipeline-summary}).

Synthesis success rates vary widely across kernels, ranging from 5.6\% to 90.7\%. The variation is driven primarily by memory interface complexity: kernels with many distinct arrays requiring simultaneous access (e.g., \texttt{PairLJCut} accesses over a dozen arrays including position, force, neighbor lists, and coefficient tables, each requiring independent memory ports under unrolling or pipelining) have a larger fraction of infeasible configurations, while kernels with fewer arrays and more regular access patterns achieve higher success rates. The two QMCPACK kernels consistently achieve higher synthesis rates (75.3\% and 90.7\%) than the LAMMPS kernels, reflecting their simpler data structure dependencies and more regular memory access patterns.

\subsection{Accelerator vs.\ Edge GPU Comparison}
\label{sec:gpu-comparison}

To demonstrate the practical value of A3D-generated accelerators, we compare them against an edge GPU. A3D processed the GPU-accelerated LAMMPS workload, identified the \texttt{k\_lj\_fast()} CUDA kernel, which accelerates the Lennard-Jones computation with GPUs, and generated accelerator designs through its full pipeline. A3D successfully re-implemented the CUDA kernel into an HLS-compatible C++ kernel, thus enabling the direct integration of the generated accelerator into the original GPU-accelerated execution flow. This demonstrates A3D's ability to accelerate workloads that are already accelerated on GPUs, which is a common scenario in scientific computing where researchers often use GPU-accelerated libraries but still face bottlenecks, as the irregular compute patterns of these kernels make them a poor fit for GPU acceleration.

The generated accelerators were synthesized using the 22\,nm technology from Global Foundries. An data-driven architecture simulator was implemented to estimate latency and energy consumption of each design point. Since the Jetson AGX Orin uses Samsung's 8\,nm process, we scale the A3D power figures to 8\,nm using a conservative $0.30\times$ factor.

The results are summarized in Table~\ref{tab:gpu-comparison}. The A3D low-latency design achieves a power consumption of 1.04\,W---a $3.5\times$ power advantage over the Jetson's 3.6\,W---with a comparable energy per execution (8.2\,mJ vs.\ 7.6\,mJ) and an EDP of 64.1\,$\mu$Js compared to the Jetson's 15.9\,$\mu$Js. The remaining latency and EDP gap is driven by the Jetson's higher degree of parallelism. The fastest A3D design occupies only 49,494\,$\mu$m$^2$, which is a small fraction of the area of the Orin SoC. Crucially, these accelerator designs were generated with \emph{zero human intervention} from a CUDA kernel, which is a novel capability.

\begin{table}[t]
\centering
\caption{Comparison of A3D-generated accelerator designs against the NVIDIA Jetson AGX Orin for the \texttt{PairLJCut} kernel. A3D power and EDP are scaled from 22\,nm to 8\,nm ($0.30\times$ power factor) for an iso-technology comparison.}
\label{tab:gpu-comparison}
\small
\begin{tabular}{lccc}
    \toprule
    \textbf{Platform} &
    \textbf{Latency} &
    \textbf{Power} &
    \textbf{EDP} \\
    \midrule
    Jetson AGX Orin           & 2.1\,ms  & 3.6\,W  & 15.9\,$\mu$Js \\
    A3D Design (low latency)  & 7.85\,ms & 1.04\,W & 64.1\,$\mu$Js \\
    A3D Design (low power)    & 618\,ms  & 0.87\,W & 0.33\,Js \\
    \bottomrule
\end{tabular}
\end{table}

\subsection{Ablation Study: Impact of Agentic Scaffolding}
\label{sec:ablation}

A central claim of A3D is that its multi-layered agentic architecture, comprising task decomposition into specialist agents, augmentation with custom tools, and adversarial verification loops, is essential for reliable agentic hardware generation. To rigorously validate this claim, we conduct an ablation study on the \texttt{PairLJCut::Compute()} kernel from LAMMPS, measuring the success rate of the HLS preparation phase under four progressively richer agent design configurations.

\subsubsection{Experimental Design}

We define four ablation configurations, each adding one architectural layer atop the previous:

\begin{itemize}
    \item \textbf{Config A --- Monolithic Agent:} A single monolithic HLS Preparer agent performs all preparation tasks without task decomposition.
    \item \textbf{Config B --- Multi-Agent:} The preparation task is decomposed into six sequential phases, each handled by a dedicated specialist agent coordinated by an orchestrator: (0)~datastructure pruning, (1)~memory interface refactoring, (2)~type conversion to \texttt{ac\_types}, (3)~loop labeling for HLS directives, (4)~math function replacement with \texttt{ac\_math} equivalents, and (5)~construct verification and cleanup.
    \item \textbf{Config C --- Agents + AST Tools:} Along with the multi-agent decomposition, we integrate AST-based tools for deterministic code transformations to augment the agents' capabilities. Agent 1 invokes the \texttt{staticMem} tool for kernel interface refactoring and Agent 2 invokes the \texttt{literalTypecast} tool for systematic typecasting of numeric literals.
    \item \textbf{Config D --- Agents + Tools + Verifiers:} The full A3D architecture comprising of decomposed specialist agents augmented with AST tools along with an adversarial verification loop for each specialist phase. After each specialist agent completes its task, a dedicated verifier agent performs targeted checks to identify any errors or omissions in the expected deliverables of that phase, providing feedback for iterative refinement until the specialist's output passes all verification checks.
\end{itemize}

For each configuration, we execute $n$ independent trials using identical inputs (the isolated \texttt{PairLJCut::Compute()} test harness implemented by Harness Developer and I/O snapshots produced by the Snapshot Engineer). Each trial is validated by a three-stage short-circuit evaluation pipeline: (1)~\textit{compile}: the refactored kernel and test harness must compile without errors, (2)~\textit{execute}: the compiled harness must run and produce outputs that match the golden outputs, and (3)~\textit{synthesize}: the HLS tool must successfully synthesize the refactored kernel. A trial passes only if all three stages succeed; evaluation halts at the first failure. We model the success rate using a Beta--Binomial Bayesian framework with a uniform $\text{Beta}(1,1)$ prior, employing adaptive stopping based on posterior precision ($\pm 12.5\%$) and success ($P(\theta > 90\%) > 0.9$)/futility ($P(\theta < 10\%) > 0.33$) thresholds. We use Qwen3.5-27B~\cite{qwen3.5} for all agents in this ablation study to isolate the impact of architectural design choices from model capabilities. We note that Qwen3.5-27B is a state-of-the-art LLM with strong reasoning and code generation capabilities scoring 72.4\% on SWE-bench-Verified~\cite{jimenez2024swe} vs. Claude Sonnet 4.5's 71.4\%.

\subsubsection{Success Rate Analysis}
\begin{figure}[t]
    \centering

\newcommand{\postA}{32.5}  
\newcommand{\ciLoA}{22.5}  
\newcommand{\ciHiA}{43.3}  
\newcommand{\nA}{75}       
\newcommand{\kA}{24}       

\newcommand{\postB}{57.3}  
\newcommand{\ciLoB}{46.1}  
\newcommand{\ciHiB}{68.2}  
\newcommand{\nB}{73}       
\newcommand{\kB}{42}       

\newcommand{\postC}{74.2}  
\newcommand{\ciLoC}{62.7}  
\newcommand{\ciHiC}{84.2}  
\newcommand{\nC}{60}       
\newcommand{\kC}{45}       

\newcommand{\postD}{93.8}  
\newcommand{\ciLoD}{83.3}  
\newcommand{\ciHiD}{99.2}  
\newcommand{\nD}{30}       
\newcommand{\kD}{29}       

\begin{tikzpicture}[
    >=stealth,
    font=\sffamily
]

\definecolor{configA}{RGB}{189, 57, 60}   
\definecolor{configB}{RGB}{217, 142, 57}  
\definecolor{configC}{RGB}{86, 152, 195}  
\definecolor{configD}{RGB}{55, 135, 75}   

\def\barwidth{1.4}      
\def\bargap{0.6}        
\def\maxheight{4.5}     
\def\scalefactor{0.045} 

\pgfmathsetmacro{\xA}{0}
\pgfmathsetmacro{\xB}{\xA + \barwidth + \bargap}
\pgfmathsetmacro{\xC}{\xB + \barwidth + \bargap}
\pgfmathsetmacro{\xD}{\xC + \barwidth + \bargap}

\draw[thick, ->] (-0.3, 0) -- (-0.3, \maxheight + 0.5) node[above, font=\small\bfseries] {Success Rate (\%)};
\draw[thick] (-0.3, 0) -- (\xD + \barwidth + 0.3, 0);

\foreach \pct in {0, 20, 40, 60, 80, 100} {
    \pgfmathsetmacro{\ypos}{\pct * \scalefactor}
    \draw[thick] (-0.3, \ypos) -- (-0.5, \ypos) node[left, font=\small] {\pct};
}

\pgfmathsetmacro{\rawA}{\kA/\nA * 100 * \scalefactor}
\pgfmathsetmacro{\postHtA}{\postA * \scalefactor}
\pgfmathsetmacro{\errLoA}{\ciLoA * \scalefactor}
\pgfmathsetmacro{\errHiA}{\ciHiA * \scalefactor}
\fill[configA, rounded corners=1pt] (\xA, 0) rectangle (\xA + \barwidth, \rawA);
\draw[black, thick, dashed] (\xA + 0.1, \postHtA) -- (\xA + \barwidth - 0.1, \postHtA);
\draw[black, very thick] (\xA + \barwidth/2, \errLoA) -- (\xA + \barwidth/2, \errHiA);
\draw[black, very thick] (\xA + \barwidth/2 - 0.15, \errLoA) -- (\xA + \barwidth/2 + 0.15, \errLoA);
\draw[black, very thick] (\xA + \barwidth/2 - 0.15, \errHiA) -- (\xA + \barwidth/2 + 0.15, \errHiA);
\node[below, font=\small, text width=1.6cm, align=center] at (\xA + \barwidth/2, -0.1) {(A)\\Monolithic\\Agent};
\node[above, font=\scriptsize] at (\xA + \barwidth/2, \errHiA + 0.1) {\kA/\nA};

\pgfmathsetmacro{\rawB}{\kB/\nB * 100 * \scalefactor}
\pgfmathsetmacro{\postHtB}{\postB * \scalefactor}
\pgfmathsetmacro{\errLoB}{\ciLoB * \scalefactor}
\pgfmathsetmacro{\errHiB}{\ciHiB * \scalefactor}
\fill[configB, rounded corners=1pt] (\xB, 0) rectangle (\xB + \barwidth, \rawB);
\draw[black, thick, dashed] (\xB + 0.1, \postHtB) -- (\xB + \barwidth - 0.1, \postHtB);
\draw[black, very thick] (\xB + \barwidth/2, \errLoB) -- (\xB + \barwidth/2, \errHiB);
\draw[black, very thick] (\xB + \barwidth/2 - 0.15, \errLoB) -- (\xB + \barwidth/2 + 0.15, \errLoB);
\draw[black, very thick] (\xB + \barwidth/2 - 0.15, \errHiB) -- (\xB + \barwidth/2 + 0.15, \errHiB);
\node[below, font=\small, text width=1.6cm, align=center] at (\xB + \barwidth/2, -0.1) {(B)\\Multi-\\Agent};
\node[above, font=\scriptsize] at (\xB + \barwidth/2, \errHiB + 0.1) {\kB/\nB};

\pgfmathsetmacro{\rawC}{\kC/\nC * 100 * \scalefactor}
\pgfmathsetmacro{\postHtC}{\postC * \scalefactor}
\pgfmathsetmacro{\errLoC}{\ciLoC * \scalefactor}
\pgfmathsetmacro{\errHiC}{\ciHiC * \scalefactor}
\fill[configC, rounded corners=1pt] (\xC, 0) rectangle (\xC + \barwidth, \rawC);
\draw[black, thick, dashed] (\xC + 0.1, \postHtC) -- (\xC + \barwidth - 0.1, \postHtC);
\draw[black, very thick] (\xC + \barwidth/2, \errLoC) -- (\xC + \barwidth/2, \errHiC);
\draw[black, very thick] (\xC + \barwidth/2 - 0.15, \errLoC) -- (\xC + \barwidth/2 + 0.15, \errLoC);
\draw[black, very thick] (\xC + \barwidth/2 - 0.15, \errHiC) -- (\xC + \barwidth/2 + 0.15, \errHiC);
\node[below, font=\small, text width=1.8cm, align=center] at (\xC + \barwidth/2, -0.1) {(C)\\+ AST\\Tools};
\node[above, font=\scriptsize] at (\xC + \barwidth/2, \errHiC + 0.1) {\kC/\nC};

\pgfmathsetmacro{\rawD}{\kD/\nD * 100 * \scalefactor}
\pgfmathsetmacro{\postHtD}{\postD * \scalefactor}
\pgfmathsetmacro{\errLoD}{\ciLoD * \scalefactor}
\pgfmathsetmacro{\errHiD}{\ciHiD * \scalefactor}
\fill[configD, rounded corners=1pt] (\xD, 0) rectangle (\xD + \barwidth, \rawD);
\draw[black, thick, dashed] (\xD + 0.1, \postHtD) -- (\xD + \barwidth - 0.1, \postHtD);
\draw[black, very thick] (\xD + \barwidth/2, \errLoD) -- (\xD + \barwidth/2, \errHiD);
\draw[black, very thick] (\xD + \barwidth/2 - 0.15, \errLoD) -- (\xD + \barwidth/2 + 0.15, \errLoD);
\draw[black, very thick] (\xD + \barwidth/2 - 0.15, \errHiD) -- (\xD + \barwidth/2 + 0.15, \errHiD);
\node[below, font=\small, text width=1.6cm, align=center] at (\xD + \barwidth/2, -0.1) {(D)\\+ Verifier\\Loops};
\node[above, font=\scriptsize] at (\xD + \barwidth/2, \errHiD + 0.1) {\kD/\nD};

\pgfmathsetmacro{\arrowY}{\maxheight + 0.15}
\draw[->, thick, configB] (\xA + \barwidth + 0.05, \arrowY) -- (\xB - 0.05, \arrowY) 
    node[midway, above, font=\scriptsize\itshape, text=black] {decompose};
\draw[->, thick, configC] (\xB + \barwidth + 0.05, \arrowY) -- (\xC - 0.05, \arrowY)
    node[midway, above, font=\scriptsize\itshape, text=black] {tooling};
\draw[->, thick, configD] (\xC + \barwidth + 0.05, \arrowY) -- (\xD - 0.05, \arrowY)
    node[midway, above, font=\scriptsize\itshape, text=black] {verify};

\end{tikzpicture}
    \caption{Success rates for the HLS preparation phase under four ablation configurations. Bar heights indicate the raw pass rate ($k/n$); dashed lines mark the Bayesian posterior mean; error bars show 95\% credible intervals. Each layer of A3D's architecture---task decomposition (A$\to$B), tool augmentation (B$\to$C), and verification loops (C$\to$D)---contributes a measurable improvement in reliability.}
    \label{fig:ablation-success}
\end{figure}
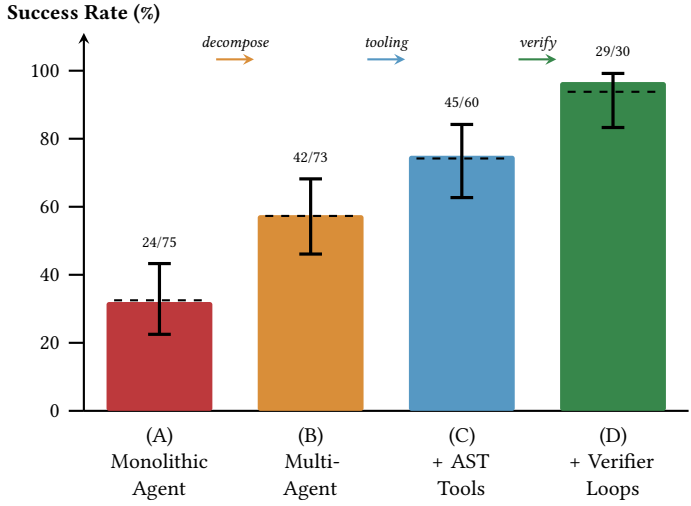

\begin{table}[t]
\centering
\caption{Bayesian posterior analysis of success rates across ablation configurations. $n$: completed trials, $k$: passed trials, CI: 95\% credible interval.}
\label{tab:ablation-bayesian}
\small
\begin{tabular}{lccccccc}
    \toprule
    \textbf{Config} & $n$ & $k$ &
    \textbf{\begin{tabular}[c]{@{}c@{}}Pass\\Rate\end{tabular}} &
    \textbf{\begin{tabular}[c]{@{}c@{}}Post.\\Mean\end{tabular}} &
    \textbf{95\% CI} &
    $P(\theta\!>\!90)$ \\
    \midrule
    (A) Monolithic   & 75 & 24 & 32.0 & 32.5 & [22.5, 43.3] & 0.0 \\
    (B) Multi-Agent   & 73 & 42 & 57.5 & 57.3 & [46.1, 68.2] & 0.0 \\
    (C) + Tools      & 60 & 45 & 75.0 & 74.2 & [62.7, 84.2] & 0.0 \\
    (D) + Verifiers  & 30 & 29 & 96.7 & 93.8 & [83.3, 99.2] & 0.83 \\
    \bottomrule
\end{tabular}
\end{table}

Figure~\ref{fig:ablation-success} presents the posterior mean success rate for each configuration with 95\% credible intervals. The results show a clear upward staircase as A3D features are added.

Table~\ref{tab:ablation-bayesian} presents the full Bayesian analysis. The posterior mean success rate increases from 32.5\% (Config~A) to 93.8\% (Config~D), with each architectural layer contributing a substantial improvement. Notably, only the full A3D architecture (Config~D) achieves a nonzero probability of exceeding a 90\% success rate ($P(\theta > 90\%) = 0.83$), while all other configurations have effectively zero probability of reaching this threshold.

Going from Config~A to Config~B demonstrates the value of \textbf{task decomposition}: the pass rate nearly doubles from 32.0\% to 57.5\%. A monolithic agent cannot manage the cognitive complexity of refactoring 35 unsupported constructs across nearly 1500 AST nodes within a single context window, frequently triggering context condensation and losing track of earlier transformations or introducing inconsistencies.
The improvement from Config~B to Config~C validates \textbf{tool augmentation}: the pass rate rises from 57.5\% to 75.0\%. While LLMs can attempt code transformations like literal typecasting and pointer-to-array conversion generatively, the AST-based tools handle these transformations deterministically and reliably, eliminating a significant source of errors.
Finally, the gain from Config~C to Config~D validates the \textbf{specialist--verifier loop}: the pass rate jumps from 75.0\% to 96.7\%. Adversarial verification catches subtle errors that specialist agents miss, such as incomplete transformations. This final layer ultimately pushes A3D into the high-reliability regime, as reflected by the $P(\theta > 90\%) = 0.83$ in Config~D versus zero for all other configurations.

\subsubsection{Cost--Efficiency Analysis}
\begin{table*}[bp!]
\centering
\caption{Cost and efficiency metrics across ablation configurations. Cost per success is computed as mean cost / pass rate.}
\label{tab:ablation-cost}
\small
\begin{tabular}{lccccccccc}
    \toprule
    \textbf{Config} &
    $n$ & $k$ &
    \textbf{\begin{tabular}[c]{@{}c@{}}Mean\\Cost (\$)\end{tabular}} &
    \textbf{\begin{tabular}[c]{@{}c@{}}Total\\Cost (\$)\end{tabular}} &
    \textbf{\begin{tabular}[c]{@{}c@{}}Mean\\Time (min)\end{tabular}} &
    \textbf{\begin{tabular}[c]{@{}c@{}}Mean\\Tokens In\end{tabular}} &
    \textbf{\begin{tabular}[c]{@{}c@{}}Mean\\Tokens Out\end{tabular}} &
    \textbf{\begin{tabular}[c]{@{}c@{}}Mean\\API Calls\end{tabular}} &
    \textbf{\begin{tabular}[c]{@{}c@{}}Cost per\\Success (\$)\end{tabular}} \\
    \midrule
    (A) Monolithic   & 75 & 24 & 2.65 & 198.75 & 12m13s & 9,311,098 & 58,166 & 86 & 9.17 \\
    (B) Multi-Agent   & 73 & 42 & 3.66 & 267.18 & 20m0s & 11,479,576 & 91,715 & 169 & 6.37 \\
    (C) + Tools      & 60 & 45 & 4.07 & 244.20 & 23m20s & 12,662,270 & 112,522 & 191 & 5.43 \\
    (D) + Verifiers  & 30 & 29 & 7.33 & 219.90 & 37m19s & 23,071,153 & 170,735 & 303 & 7.58 \\
    \bottomrule
\end{tabular}
\end{table*}

While simpler configurations are cheaper, they are less reliable. Table~\ref{tab:ablation-cost} presents the mean cost and efficiency metrics for each configuration.
Mean per-trial cost increases monotonically from \$2.65 (Config~A) to \$7.33 (Config~D), driven primarily by higher token consumption. Config~D processes 23M input tokens and 171K output tokens per trial, roughly 2.5$\times$ and 3$\times$ that of Config~A, respectively. The higher token counts reflect the overhead of multi-agent coordination, tool invocations, and verifier feedback loops.

The \textit{cost per successful outcome} (mean cost divided by pass rate) is minimized by Config~C at \$5.43, owing to its arsenal of tools. Config~D incurs a higher cost per success (\$7.58) due to verifier loop overhead. However, reliability is paramount for end-to-end automation, and Config~D is the only configuration for which $P(\theta > 90\%) > 0$, indicating that the verifier loop is necessary to achieve high-confidence automation. At a 75\% pass rate (Config~C), one in four runs requires human intervention; at 97\% (Config~D), this drops to approximately one in thirty-three, a reduction that is critical for unmanned deployment. 

We report that the observations from this ablation study guided the design of A3D's architecture based on all three principles of task decomposition, tool augmentation, and verification loops to achieve similar levels of reliability at every step of the A3D workflow.

\subsubsection{Verifier Loop Behavior}
\begin{table}[t]
\centering
\caption{Per-phase verifier loop behavior in Config~D (30 trials). Single-pass rate: fraction of episodes where the verifier accepted the specialist's output on the first attempt.}
\label{tab:verifier-episodes}
\small
\begin{tabular}{lcccc}
    \toprule
    \textbf{Phase} &
    \textbf{\begin{tabular}[c]{@{}c@{}}Mean Verifier\\Rounds\end{tabular}} &
    \textbf{\begin{tabular}[c]{@{}c@{}}Max Verifier\\Rounds\end{tabular}} &
    \textbf{\begin{tabular}[c]{@{}c@{}}Single-Pass\\Rate\end{tabular}} \\
    \midrule
    0 --- Pruning       & 1.00 & 1 & 100\% \\
    1 --- Memory        & 1.23 & 2 & 76.7\% \\
    2 --- Types         & 1.27 & 3 & 76.7\% \\
    3 --- Loop labels   & 1.00 & 1 & 100\% \\
    4 --- Math          & 1.03 & 2 & 96.7\% \\
    5 --- Verification  & 1.00 & 1 & 100\% \\
    \bottomrule
\end{tabular}
\end{table}

To understand \emph{where} the verifier loop's overhead materializes, we parse the agent execution traces from Config~D into \textit{specialist episodes}, defined as the subsequence of mode transitions between when the orchestrator dispatches to a specialist and when control returns to the orchestrator. For each of the six preparation phases, we measure the number of verifier rounds (specialist$\to$verifier transitions) per episode and the \textit{single-pass rate}: the fraction of episodes in which the specialist's output was accepted by the verifier on the first attempt.

Table~\ref{tab:verifier-episodes} presents the per-phase results across all 30 completed Config~D trials.
Phases~0 (datastructure pruning), 3 (loop labeling), and 5 (construct verification) achieve a 100\% single-pass rate: the verifier never rejects the specialist's output. Phase~4 (math function replacement) is nearly as stable at 96.7\%, with only one trial requiring a second verifier round. In contrast, Phases~1 (memory interface refactoring) and 2 (type conversion) both achieve a 76.7\% single-pass rate, with Phase~2 exhibiting the highest maximum verifier rounds (3) and transitions (6) of any phase. These phases involve the most complex, multi-site code transformations: converting multi-level pointer indirection to static arrays (Phase~1) and replacing all numeric types with HLS-compatible \texttt{ac\_types} (Phase~2).

Importantly, the per-phase fragility profile is \emph{kernel-specific}. The \texttt{PairLJCut::Compute()} kernel used in this ablation has relatively simple data structures and straightforward arithmetic. In contrast, the \texttt{ReaxFF::Torsion\_Angles()} kernel described in \S\ref{sec:e2e-flow} requires  processing far more complex data structures and transcendental math functions making the verifier crucial for phases 0 and 4 as well. These results should therefore be interpreted as characterizing the agentic flow on one representative kernel rather than as universal difficulty rankings across phases.


\subsection{End-to-End Flow: A Case Study}
\label{sec:e2e-flow}

To illustrate how A3D operates in practice, we present its end-to-end processing of the \texttt{ReaxFF::Torsion\_Angles()} kernel from LAMMPS. This kernel evaluates torsion angle energy contributions in the Reactive Force Field and is representative of the most challenging kernel in our benchmark suite.

\subsubsection{Analysis Phase}

The \textit{Workload Analyst} began by exploring the LAMMPS codebase using semantic search. It identified the \textit{reaxff} workload as exercising the reactive force field potential, mapping the execution flow from the main simulation loop through the ReaxFF force computation routines to the torsion angle evaluation kernel.

The \textit{Performance Profiler} compiled LAMMPS with profiling instrumentation and executed the \textit{reaxff} workload. The profiling results identified \texttt{ReaxFF::Torsion\_Angles()} as a dominant performance bottleneck, owing to its deeply nested loop structure. 

The \textit{Bottleneck Analyst} then used CodeQL to perform static analysis, revealing that \texttt{Torsion\_Angles()} depends on multiple complex data structures spanning several class hierarchies. The Bottleneck Analyst dynamically generates CodeQL queries tailored to each kernel it analyzes. Figure~\ref{fig:codeql-query} shows the query it formulated for \texttt{Torsion\_Angles()}. The query identifies every variable accessed by the kernel, both directly and through functions it calls (one level of call depth), and classifies each access as \texttt{INPUT} or \texttt{OUTPUT}. This produces a complete I/O footprint: variable name, type, and data-flow direction. For \texttt{Torsion\_Angles()}, the query identified 25 array parameters spanning 7 struct types across 4 class hierarchy levels, producing the dependency map that guides the Snapshot Engineer (which variables to capture) and the Harness Developer (which data structures the isolated test harness must replicate). Figure~\ref{fig:torsion-structs} shows a subset of these data structures as defined in the original LAMMPS source (\texttt{reaxff\_types.h}). The Bottleneck Analyst then performs dynamic analysis to determine the runtime sizes and behavior of these data structures with the representative workload.

\begin{figure}[t]
\begin{lstlisting}[language=SQL, basicstyle=\ttfamily\small, morekeywords={from,where,select,import,exists,if,then,else,and,or,not}, deletekeywords={type}]
import cpp

from Function f, Variable v,
     VariableAccess va, string usage,
     Function calledFunc
where
  f.getName() = "Torsion_Angles" and
  f.getFile().getBaseName() =
    "reaxff_torsion_angles.cpp" and
  (
    (va.getEnclosingFunction() = f) or
    (exists(FunctionCall fc |
      fc.getEnclosingFunction() = f and
      fc.getTarget() = calledFunc and
      va.getEnclosingFunction() = calledFunc
    ))
  ) and
  va.getTarget() = v and
  if va.isUsedAsLValue()
  then usage = "OUTPUT"
  else usage = "INPUT"
select va,
  v.getName() + "," +
  v.getType().toString() + "," + usage
\end{lstlisting}
\caption{CodeQL query dynamically generated by the Bottleneck Analyst to extract the I/O data-structure footprint of \texttt{Torsion\_Angles()}. The query traverses one level of call depth to capture variables accessed by helper functions.}
\label{fig:codeql-query}
\end{figure}

\begin{figure}[t]
\noindent
\begin{minipage}[t]{0.48\columnwidth}
\begin{lstlisting}[xleftmargin=0pt,xrightmargin=0pt,basicstyle=\ttfamily\scriptsize]
// Primitive types
using rvec = double[3];
using ivec = int[3];

// Atom data
struct reax_atom {
  rc_tagint orig_id;
  int type; char name[8];
  rvec x, v, f; double q;
  int Hindex, num_bonds,
      num_hbonds;
};

// Force field parameters
struct global_parameters {
  int n_global, vdw_type;
  double *l;
};
struct four_body_parameters {
  double V1, V2, V3;
  double p_tor1, p_cot1;
};
struct four_body_header {
  int cnt;
  four_body_parameters
    prm[REAX_MAX_4BODY_PARAM];
};
struct reax_interaction {
  int num_atom_types;
  global_parameters gp;
  single_body_parameters *sbp;
  two_body_parameters **tbp;
  three_body_header ***thbp;
  hbond_parameters ***hbp;
  four_body_header ****fbp;
};

// 1st arg: system
struct reax_system {
  reax_interaction reax_param;
  int n, N, numH;
  int local_cap, total_cap, Hcap;
  reax_atom *my_atoms;
  int my_bonds;
  double safezone, saferzone;
  LR_lookup_table **LR;
};

// 2nd arg: control
struct control_params {
  int nthreads;
  double bond_cut;
  double nonb_cut, nonb_low;
  double hbond_cut;
  double bg_cut, bo_cut;
  double thb_cut, thb_cutsq;
  int tabulate;
  int lgflag, enobondsflag;
};

// 3rd arg: simulation data
struct energy_data {
  double e_bond, e_ov, e_un;
  double e_lp, e_ang, e_pen;
  double e_coa, e_hb;
  double e_tor, e_con;
  double e_vdW, e_ele, e_pol;
};
\end{lstlisting}
\end{minipage}\hfill
\begin{minipage}[t]{0.48\columnwidth}
  \begin{lstlisting}[xleftmargin=0pt,xrightmargin=0pt,basicstyle=\ttfamily\scriptsize]
struct simulation_data {
  rc_bigint step;
  energy_data my_en;
};

// 4th arg: workspace storage
struct storage {
  int allocated;
  double *total_bond_order;
  double *Deltap, *Deltap_boc;
  double *Delta, *Delta_lp;
  double *Delta_lp_temp;
  double *Delta_e, *Delta_boc;
  double *Delta_val;
  double *dDelta_lp;
  double *dDelta_lp_temp;
  double *nlp, *nlp_temp;
  double *Clp, *vlpex;
  rvec *dDeltap_self;
  int *bond_mark;
  double Tap[8];
  double *CdDelta;
  rvec *f;
  rvec *forceReduction;
  double *CdDeltaReduction;
  int *valence_angle_atom_myoffset;
  double *s;
  reallocate_data realloc;
};

// 5th arg: reax_list **
struct bond_order_data {
  double BO, BO_s, BO_pi, BO_pi2;
  double Cdbo, Cdbopi, Cdbopi2;
  double C1dbo, C2dbo, C3dbo;
  double C1dbopi, C2dbopi;
  double C3dbopi, C4dbopi;
  double C1dbopi2, C2dbopi2;
  double C3dbopi2, C4dbopi2;
  rvec dBOp, dln_BOp_s;
  rvec dln_BOp_pi, dln_BOp_pi2;
  double *CdboReduction;
};

struct bond_data {
  int nbr, sym_index;
  int dbond_index;
  ivec rel_box;
  double d; rvec dvec;
  bond_order_data bo_data;
};
struct three_body_interaction_data {
  int thb, pthb;
  double theta, cos_theta;
  rvec dcos_di, dcos_dj, dcos_dk;
};
union list_type {
  three_body_interaction_data
    *three_body_list;
  bond_data *bond_list;
  far_neighbor_data *far_nbr_list;
  hbond_data *hbond_list;
};
struct reax_list {
  int allocated, n, num_intrs;
  int *index, *end_index;
  int type; list_type select;
};
\end{lstlisting}
\end{minipage}
\caption{A subset of the data structures involved in \texttt{extern void Torsion\_Angles(reax\_system*, control\_params*, simulation\_data*, storage*, reax\_list**)}, as discovered by the Bottleneck Analyst. The five function arguments expose a deep type hierarchy featuring dynamically allocated arrays, multi-level pointer indirection (\texttt{four\_body\_header~****fbp}), a union-based polymorphic list container (\texttt{list\_type}), and nested structs.}
\label{fig:torsion-structs}
\end{figure}

\subsubsection{Preparation Phase}

The \textit{Snapshot Engineer} instrumented code at the kernel call site to capture the complete input and output state of \texttt{Torsion\_Angles()} at the function boundary during workload execution. The captured input snapshot (8.9\,MB) includes atom positions, bond and three-body interaction lists (40,320 bonds, 68,659 three-body interactions), workspace arrays, and force field parameters; the output snapshot (951\,KB) captures all modified state. 

The \textit{Harness Developer} created an isolated test harness comprising three files (\texttt{bottleneck.cpp}, \texttt{bottleneck.h}, \texttt{main.cpp}) that reads the captured snapshots, invokes the extracted kernel, and validates outputs against the golden reference from the output snapshot. The harness was iteratively refined in collaboration with the Snapshot Engineer until numerical equivalence was achieved. 
The \textit{HLS Preparer} then performed the most challenging step: refactoring the kernel's unsupported constructs, such as multi-level pointer indirection, dynamic memory allocation, standard library math functions and floating point numbers, with HLS-compatible alternatives. This was orchestrated across six specialist phases, each followed by adversarial verification:

\begin{enumerate}
    \setcounter{enumi}{-1}
    \item \textbf{Datastructure Pruning:} Removed 28 unused fields across 7 structs and converted a \texttt{union} to a \texttt{struct}, thereby simplifying the kernel's type system.
    \item \textbf{Memory Interface Refactoring:} Multi-level pointer indirection in bond lists, three-body lists, and atom position arrays was converted to C-style array notation with compile-time constants, using the AST-based \texttt{staticMem} tool. The tool applied 105 automated transformations (90 in the kernel, 15 in the header). Array dimensions were determined from the captured snapshots and sized 10$\times$ larger to accommodate production workloads (e.g., 168 atoms $\to$ 2,048 static array, 40,320 bonds $\to$ 400,000). A subsequent verifier pass identified a residual \texttt{LAMMPS::Pair*} pointer that was missed; the specialist converted it to a statically-embedded struct, achieving zero remaining pointers across all 15 data structures.
    \item \textbf{Type Conversion:} Standard numeric types were replaced with HLS-compatible \texttt{ac\_types} using compile guards. The agent queried the Catapult documentation RAG for appropriate type mappings (Figure~\ref{fig:hls_preparer_example}). Nine category-specific typedefs were introduced (\texttt{Energy\_t}, \texttt{Angle\_t}, \texttt{BondOrder\_t}, \texttt{Distance\_t}, \texttt{Param\_t}, \texttt{Calc\_t}, \texttt{AtomDeriv\_t}, \texttt{BondDeriv\_t}, \texttt{Real\_t}) to enable independent precision control in the numerics exploration phase. All numeric literals were explicitly typecast using the \texttt{literalTypecast} tool to prevent implicit conversion ambiguities.
    \item \textbf{Loop Labeling:} All four nested loop structures were labeled with unique, descriptive identifiers
    to enable targeted application of HLS optimization directives in the synthesis phase.
    \item \textbf{Math Function Replacement:} Five standard math functions (\texttt{sin}, \texttt{cos}, \texttt{sqrt}, \texttt{exp}, \texttt{atan2}) were replaced with HLS-compatible \texttt{ac\_math} equivalents, consulting both local documentation and the Catapult RAG pipeline. 
    \item \textbf{Construct Verification:} A final verification pass ensured no remaining unsupported programming constructs, validated compilation under HLS flags, and confirmed numerical equivalence against the golden snapshots.
\end{enumerate}


\subsubsection{Synthesis Phase}
In the synthesis phase, the numerics of data structures and computations were optimized for hardware efficiency while preserving functional correctness, before invoking Catapult HLS for RTL generation. This phase is critical for achieving a performant hardware design, as naive type choices can lead to excessive resource usage or failed synthesis.

The \textit{Numerics Explorer} optimized number formats and bit-widths to maximize hardware efficiency while preserving functional correctness. The kernel's nine floating-point typedefs, initially set to \texttt{ac\_ieee\_float64} (IEEE-754 double), were converted to fixed-point representations (\texttt{ac\_fixed}) for more efficient hardware implementation.

A key discovery during this phase was that \textit{precision is more important than range}. The initial attempt with \texttt{ac\_fixed<64,32,true>} (32 integer bits, 31 fractional) produced 10.5\% error in torsion energy due to insufficient fractional precision for accumulated values. Reducing the integer width to 20 bits while increasing fractional bits to 43 (\texttt{ac\_fixed<64,20,true>}) passed verification, since 20 integer bits ($\pm$524,288 range) suffice for all kernel values while 43 fractional bits provide the precision needed for accurate representation.

The Numerics Explorer then performed per-type bit-width optimization on each typedef independently, respecting a \textit{scalability exception} policy: typedefs used in accumulation patterns (\texttt{Energy\_t}, \texttt{AtomDeriv\_t}, \texttt{BondDeriv\_t}) were preserved at 64 bits to prevent precision loss from compounding across interactions, while optimization candidates were reduced through binary search with verification after each step:

\begin{itemize}
    \item \texttt{BondOrder\_t}: 64 $\to$ 17 bits (73.4\% reduction) --- bounded range [0, $\sim$3]
    \item \texttt{Angle\_t}: 64 $\to$ 21 bits (67.2\%) --- bounded range [$-\pi$, $\pi$]
    \item \texttt{Param\_t}: 64 $\to$ 22 bits (65.6\%) --- constant force field parameters
    \item \texttt{Calc\_t}: 64 $\to$ 45 bits (29.7\%) --- intermediate calculations
\end{itemize}

Two typedefs (\texttt{Distance\_t}, \texttt{Real\_t}) could not be optimized due to type interdependencies: they participate in vector operations (\texttt{rvec\_ScaledAdd}) that interface with \texttt{AtomDeriv\_t} (a scalability exception), creating a type-width coupling that the agent correctly identified and documented rather than forcing an incompatible reduction.

The \textit{Synthesizer} invoked Catapult HLS to synthesize the prepared kernel. The baseline synthesis completed successfully in 1 hour 36 minutes, generating RTL netlists in both Verilog and VHDL, along with other artifacts such as synthesis reports, resource utilization summaries, and timing analyses. 

\subsubsection{Design Space Exploration}

Design space exploration is broken into three phases: (0) knob identification, (1) strategy and configuration generation, and (2) execution. The goal is to systematically explore the optimization space of HLS directives (e.g., loop unrolling, pipelining, memory partitioning) to identify Pareto-optimal configurations that balance area and latency.

\paragraph{Phase~0 --- Knob Identification.}
The DSE agent conducted 8 focused RAG queries to build comprehensive understanding of Catapult HLS optimization techniques, covering design goals, clock period constraints, loop unrolling, loop pipelining, and memory interleaving. It then performed manual code analysis of the kernel's loop hierarchy and classified all loop dependencies, finding that the innermost loop exhibits weak accumulator dependencies (energy accumulations to the same location, force accumulations to different atoms), making II=1 likely achievable. Array access patterns were documented with K-values: the atom position array requires K=4 windowed reads (simultaneously accessing atoms $i$, $j$, $k$, $l$), bond lists require K=3 reads, and three-body lists require K=2 reads.

\paragraph{Phase~1 --- Strategy and Configuration Generation.}
Using the extracted HLS resource paths (25 array parameters and 4 loop paths from the baseline synthesis log) and the knob identification results, the agent generated a YAML design space specification. Figure~\ref{fig:dse-yaml} shows an excerpt of this specification, illustrating its declarative structure. Each dimension defines its options with an HLS directive type, target resource path, and range of values. Memory interleaving factors were computed based on number of accesses per iteration. Multiple interleaving levels were explored for each unroll/pipeline option to map the boundary between feasible and infeasible designs. The agent then invoked the \texttt{genDSE} tool, which parsed this YAML specification, performed combinatorial expansion across all possible combinations, and emitted 649 synthesis-ready Tcl directive files along with a CSV summary manifest.

\begin{figure}[t]
\begin{lstlisting}[language={}, basicstyle=\ttfamily\scriptsize, keywordstyle={}, commentstyle=\itshape\color{gray}, morekeywords={}, deletekeywords={type}]
kernel_name: "Torsion_Angles"
base_hls_tcl_file: "directives/baseline.tcl"
dimensions:
  - id: "design_goal"
    type: "DESIGN_GOAL"
    values: [area, latency]

  - id: "clock_period"
    type: "CLOCK_PERIOD"
    values: [1.0, 2.0, 3.0, 5.0, 7.0, 10.0]

  - id: "innermost_loop_unroll"
    type: "UNROLL"
    target_hls_path: "/.../LOOP_TORSION_ANGLE_L"
    values: [no, 2, 4]  # with interleaving variants

  - id: "innermost_loop_pipeline"
    type: "PIPELINE_II"
    target_hls_path: "/.../LOOP_TORSION_ANGLE_L"
    values: [none, 1, 2]  # with interleaving variants
\end{lstlisting}
\caption{Excerpt of the YAML design space specification generated by the DSE agent for \texttt{Torsion\_Angles()}. Each dimension declares its HLS directive type, target resource path, and the set of values to explore. The \texttt{genDSE} tool parses this specification and performs combinatorial expansion to emit one Tcl directive file per design point.}
\label{fig:dse-yaml}
\end{figure}

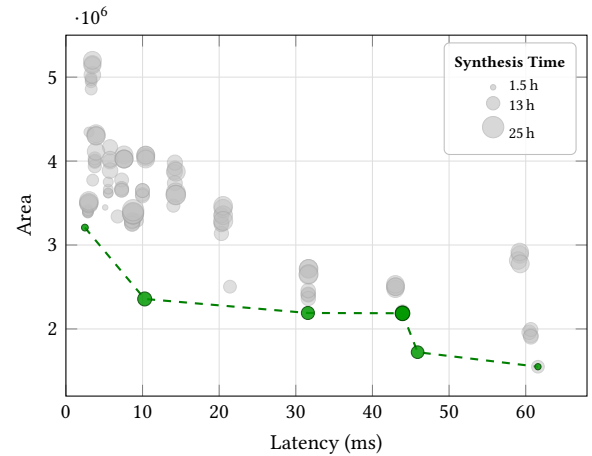
\begin{figure}[b]
\centering

\begin{tikzpicture}
\begin{axis}[
    width=\columnwidth,
    height=0.75\columnwidth,
    xlabel={Latency (ms)},
    ylabel={Area},
    xmin=0, xmax=68,
    ymin=1200000, ymax=5500000,
    grid=major,
    grid style={gray!25},
    tick label style={font=\small},
    label style={font=\small},
    scaled y ticks=true,
    yticklabel style={
        /pgf/number format/fixed,
        /pgf/number format/precision=1,
        /pgf/number format/set thousands separator={,},
    },
    clip=true,
]

\addplot[
    only marks,
    mark=*,
    scatter,
    point meta=1,
    visualization depends on={\thisrow{size_pt} \as \perpointmarksize},
    scatter/@pre marker code/.code={%
        \scope[mark size=\perpointmarksize, fill=gray!50, draw=gray!70,
               fill opacity=0.5, draw opacity=0.5]%
    },
    scatter/@post marker code/.code={\endscope},
] table[
    x=latency_ms,
    y=area_raw,
    col sep=tab,
] {sections/figs/data/torsion_dse_nonpareto.dat};

\addplot[
    no markers,
    thick,
    green!50!black,
    dashed,
] table[
    x=latency_ms,
    y=area_raw,
    col sep=tab,
] {sections/figs/data/torsion_dse_pareto.dat};

\addplot[
    only marks,
    mark=*,
    scatter,
    point meta=1,
    visualization depends on={\thisrow{size_pt} \as \perpointmarksize},
    scatter/@pre marker code/.code={%
        \scope[mark size=\perpointmarksize, fill=green!60!black, draw=green!30!black,
               fill opacity=0.85, draw opacity=0.9]%
    },
    scatter/@post marker code/.code={\endscope},
] table[
    x=latency_ms,
    y=area_raw,
    col sep=tab,
] {sections/figs/data/torsion_dse_pareto.dat};

\node[anchor=north east, font=\scriptsize, inner sep=4pt,
      fill=white, fill opacity=0.85, text opacity=1,
      draw=gray!50, rounded corners=2pt]
    at (rel axis cs:0.98,0.98) {%
    \begin{tabular}{@{}c@{}}
        \textbf{Synthesis Time} \\[2pt]
        \begin{tabular}{@{}c@{\;}l@{}}
            \tikz\fill[gray!50, draw=gray!70, opacity=0.6] (0,0) circle (1.0pt); & 1.5\,h \\
            \tikz\fill[gray!50, draw=gray!70, opacity=0.6] (0,0) circle (2.5pt); & 13\,h \\
            \tikz\fill[gray!50, draw=gray!70, opacity=0.6] (0,0) circle (4.0pt); & 25\,h \\
        \end{tabular}
    \end{tabular}%
};

\end{axis}
\end{tikzpicture}
\caption{Design space exploration results for \texttt{Torsion\_Angles()}. Each point represents a successfully synthesized configuration; marker size encodes synthesis time (1.5--25\,h). Green markers indicate the 7 Pareto-optimal designs spanning a $25\times$ latency--$2.1\times$ area tradeoff.}
\label{fig:torsion-dse}
\end{figure}

\paragraph{Phase~2 --- Parallel Execution.}
The agent invoked the \texttt{runDSE} execution tool, which scheduled the 649 configurations for parallel synthesis across 128 concurrent jobs distributed over local and remote compute nodes. The tool manages Catapult invocation with the appropriate directive file for each design point, handles remote job scheduling via SSH, and collects per-configuration synthesis logs. The total runtime for all 649 syntheses was 30 hours 27 minutes. Of the 649 configurations, 109 synthesized successfully (16.8\% success rate). The 540 failures are expected and informative: they represent under-interleaved configurations that cannot meet scheduling constraints, validating the memory pressure model from Phase~0. Configurations with position array interleaving below K=4 systematically fail, confirming the windowed access analysis. Successful configurations span the full optimization space, with area-optimized designs (slow clock, lower unroll) succeeding with minimal interleaving and latency-optimized designs (fast clock, higher unroll/pipeline) requiring full interleaving. Individual synthesis times ranged from $\sim$1.5 hours for simple configurations to up to 25 hours for complex designs.

Figure~\ref{fig:torsion-dse} shows the complete design space for \texttt{Torsion\_Angles}, plotting latency against area for all 109 successfully synthesized configurations. The Pareto front identifies 7 non-dominated designs spanning a $25\times$ latency range (2.5--61.6\,ms) and a $2.1\times$ area range.

\section{Conclusion}
\label{sec:conclusion}

We present A3D, the first agentic AI system for end-to-end automation of hardware design from application codebases to diverse accelerator architectures. Starting from a C/C++ or CUDA codebase and a representative workload, A3D autonomously profiles the application, identifies performance bottlenecks, extracts and isolates target kernels, refactors them for HLS compatibility, performs high-level synthesis, and explores the design space, all with zero human intervention. We demonstrated A3D on two production-scale scientific computing applications, processing kernels that span a wide range of HLS challenges. An ablation study validated A3D's three-layer architecture: task decomposition, deterministic tool augmentation, and adversarial specialist--verifier loops each contribute measurably to reliability, and only their combination achieves the high success rates required for unmanned deployment. These findings suggest that reliable agentic automation of complex engineering tasks requires the synergy of all three principles.

\bibliographystyle{ACM-Reference-Format}
\bibliography{a3d}

@article{zhao2023survey,
  title={A survey of large language models},
  author={Zhao, Wayne Xin and Zhou, Kun and Li, Junyi and Tang, Tianyi and Wang, Xiaolei and Hou, Yupeng and Min, Yingqian and Zhang, Beichen and Zhang, Junjie and Dong, Zican and others},
  journal={arXiv preprint arXiv:2303.18223},
  volume={1},
  number={2},
  pages={1--124},
  year={2023}
}

@article{bie2024renaissance,
  title={Renaissance: A survey into ai text-to-image generation in the era of large model},
  author={Bie, Fengxiang and Yang, Yibo and Zhou, Zhongzhu and Ghanem, Adam and Zhang, Minjia and Yao, Zhewei and Wu, Xiaoxia and Holmes, Connor and Golnari, Pareesa and Clifton, David A and others},
  journal={IEEE transactions on pattern analysis and machine intelligence},
  volume={47},
  number={3},
  pages={2212--2231},
  year={2024},
  publisher={IEEE}
}

@article{gao2025seedance,
  title={Seedance 1.0: Exploring the boundaries of video generation models},
  author={Gao, Yu and Guo, Haoyuan and Hoang, Tuyen and Huang, Weilin and Jiang, Lu and Kong, Fangyuan and Li, Huixia and Li, Jiashi and Li, Liang and Li, Xiaojie and others},
  journal={arXiv preprint arXiv:2506.09113},
  year={2025}
}

@article{team2025kling,
  title={Kling-Omni Technical Report},
  author={Team, Kling and Chen, Jialu and Ci, Yuanzheng and Du, Xiangyu and Feng, Zipeng and Gai, Kun and Guo, Sainan and Han, Feng and He, Jingbin and He, Kang and others},
  journal={arXiv preprint arXiv:2512.16776},
  year={2025}
}

@article{jiang2026survey,
  title={A survey on large language models for code generation},
  author={Jiang, Juyong and Wang, Fan and Shen, Jiasi and Kim, Sungju and Kim, Sunghun},
  journal={ACM Transactions on Software Engineering and Methodology},
  volume={35},
  number={2},
  pages={1--72},
  year={2026},
  publisher={ACM New York, NY}
}

@misc{novikov2025alphaevolvecodingagentscientific,
      title={AlphaEvolve: A coding agent for scientific and algorithmic discovery}, 
      author={Alexander Novikov and Ngân Vũ and Marvin Eisenberger and Emilien Dupont and Po-Sen Huang and Adam Zsolt Wagner and Sergey Shirobokov and Borislav Kozlovskii and Francisco J. R. Ruiz and Abbas Mehrabian and M. Pawan Kumar and Abigail See and Swarat Chaudhuri and George Holland and Alex Davies and Sebastian Nowozin and Pushmeet Kohli and Matej Balog},
      year={2025},
      eprint={2506.13131},
      archivePrefix={arXiv},
      primaryClass={cs.AI},
      url={https://arxiv.org/abs/2506.13131}, 
}

@inproceedings{bai2024longbench,
  title={Longbench: A bilingual, multitask benchmark for long context understanding},
  author={Bai, Yushi and Lv, Xin and Zhang, Jiajie and Lyu, Hongchang and Tang, Jiankai and Huang, Zhidian and Du, Zhengxiao and Liu, Xiao and Zeng, Aohan and Hou, Lei and others},
  booktitle={Proceedings of the 62nd annual meeting of the association for computational linguistics (volume 1: Long papers)},
  pages={3119--3137},
  year={2024}
}

@article{chien201110x10,
  title={10x10: A general-purpose architectural approach to heterogeneity and energy efficiency},
  author={Chien, Andrew A and Snavely, Allan and Gahagan, Mark},
  journal={Procedia Computer Science},
  volume={4},
  pages={1987--1996},
  year={2011},
  publisher={Elsevier}
}

@article{chen2016eyeriss,
  title={Eyeriss: An energy-efficient reconfigurable accelerator for deep convolutional neural networks},
  author={Chen, Yu-Hsin and Krishna, Tushar and Emer, Joel S and Sze, Vivienne},
  journal={IEEE journal of solid-state circuits},
  volume={52},
  number={1},
  pages={127--138},
  year={2016},
  publisher={IEEE}
}

@inproceedings{jouppi2017datacenter,
  title={In-datacenter performance analysis of a tensor processing unit},
  author={Jouppi, Norman P and Young, Cliff and Patil, Nishant and Patterson, David and Agrawal, Gaurav and Bajwa, Raminder and Bates, Sarah and Bhatia, Suresh and Boden, Nan and Borchers, Al and others},
  booktitle={Proceedings of the 44th annual international symposium on computer architecture},
  pages={1--12},
  year={2017}
}

@inproceedings{samardzic2021f1,
  title={F1: A fast and programmable accelerator for fully homomorphic encryption},
  author={Samardzic, Nikola and Feldmann, Axel and Krastev, Aleksandar and Devadas, Srinivas and Dreslinski, Ronald and Peikert, Christopher and Sanchez, Daniel},
  booktitle={MICRO-54: 54th Annual IEEE/ACM International Symposium on Microarchitecture},
  pages={238--252},
  year={2021}
}

@article{hennessy2019new,
  title={A new golden age for computer architecture},
  author={Hennessy, John L and Patterson, David A},
  journal={Communications of the ACM},
  volume={62},
  number={2},
  pages={48--60},
  year={2019},
  publisher={ACM New York, NY, USA}
}

@article{lahti2018we,
  title={Are we there yet? A study on the state of high-level synthesis},
  author={Lahti, Sakari and Sj{\"o}vall, Panu and Vanne, Jarno and H{\"a}m{\"a}l{\"a}inen, Timo D},
  journal={IEEE Transactions on Computer-Aided Design of Integrated Circuits and Systems},
  volume={38},
  number={5},
  pages={898--911},
  year={2018},
  publisher={IEEE}
}

@article{martin2009high,
  title={High-level synthesis: Past, present, and future},
  author={Martin, Grant and Smith, Gary},
  journal={IEEE Design \& Test of Computers},
  volume={26},
  number={4},
  pages={18--25},
  year={2009},
  publisher={IEEE}
}

@inproceedings{reagen2014machsuite,
  title={MachSuite: Benchmarks for accelerator design and customized architectures},
  author={Reagen, Brandon and Adolf, Robert and Shao, Yakun Sophia and Wei, Gu-Yeon and Brooks, David},
  booktitle={2014 IEEE International Symposium on Workload Characterization (IISWC)},
  pages={110--119},
  year={2014},
  organization={IEEE}
}

@inproceedings{hara2008chstone,
  title={Chstone: A benchmark program suite for practical c-based high-level synthesis},
  author={Hara, Yuko and Tomiyama, Hiroyuki and Honda, Shinya and Takada, Hiroaki and Ishii, Katsuya},
  booktitle={2008 IEEE International Symposium on Circuits and Systems (ISCAS)},
  pages={1192--1195},
  year={2008},
  organization={IEEE}
}

@inproceedings{wang2025hlsdebugger,
  title={HLSDebugger: Identification and Correction of Logic Bugs in HLS Code with LLM Solutions},
  author={Wang, Jing and Liu, Shang and Lu, Yao and Xie, Zhiyao},
  booktitle={2025 IEEE/ACM International Conference On Computer Aided Design (ICCAD)},
  pages={1--9},
  year={2025},
  organization={IEEE}
}

@article{oztas2024agentic,
  title={Agentic-HLS: An agentic reasoning based high-level synthesis system using large language models (AI for EDA workshop 2024)},
  author={Oztas, Ali Emre and Jelodari, Mahdi},
  journal={arXiv preprint arXiv:2412.01604},
  year={2024}
}

@article{yao2022react,
  title={React: Synergizing reasoning and acting in language models},
  author={Yao, Shunyu and Zhao, Jeffrey and Yu, Dian and Du, Nan and Shafran, Izhak and Narasimhan, Karthik and Cao, Yuan},
  journal={arXiv preprint arXiv:2210.03629},
  year={2022}
}

@article{huang2025survey,
  title={A survey on hallucination in large language models: Principles, taxonomy, challenges, and open questions},
  author={Huang, Lei and Yu, Weijiang and Ma, Weitao and Zhong, Weihong and Feng, Zhangyin and Wang, Haotian and Chen, Qianglong and Peng, Weihua and Feng, Xiaocheng and Qin, Bing and others},
  journal={ACM Transactions on Information Systems},
  volume={43},
  number={2},
  pages={1--55},
  year={2025},
  publisher={ACM New York, NY}
}

@article{zhang2025qwen3,
  title={Qwen3 embedding: Advancing text embedding and reranking through foundation models},
  author={Zhang, Yanzhao and Li, Mingxin and Long, Dingkun and Zhang, Xin and Lin, Huan and Yang, Baosong and Xie, Pengjun and Yang, An and Liu, Dayiheng and Lin, Junyang and others},
  journal={arXiv preprint arXiv:2506.05176},
  year={2025}
}

@misc{qwen3.5,
    title  = {{Qwen3.5}: Towards Native Multimodal Agents},
    author = {{Qwen Team}},
    month  = {February},
    year   = {2026},
    url    = {https://qwen.ai/blog?id=qwen3.5}
}

@inproceedings{jimenez2024swe,
  title={Swe-bench: Can language models resolve real-world github issues?},
  author={Jimenez, Carlos E and Yang, John and Wettig, Alexander and Yao, Shunyu and Pei, Kexin and Press, Ofir and Narasimhan, Karthik},
  booktitle={International Conference on Learning Representations},
  volume={2024},
  pages={54107--54157},
  year={2024}
}

@article{schafer2019high,
  title={High-level synthesis design space exploration: Past, present, and future},
  author={Schafer, Benjamin Carrion and Wang, Zi},
  journal={IEEE Transactions on Computer-Aided Design of Integrated Circuits and Systems},
  volume={39},
  number={10},
  pages={2628--2639},
  year={2019},
  publisher={IEEE}
}

@article{sohrabizadeh2022autodse,
  title={AutoDSE: Enabling software programmers to design efficient FPGA accelerators},
  author={Sohrabizadeh, Atefeh and Yu, Cody Hao and Gao, Min and Cong, Jason},
  journal={ACM Transactions on Design Automation of Electronic Systems (TODAES)},
  volume={27},
  number={4},
  pages={1--27},
  year={2022},
  publisher={ACM New York, NY}
}

@misc{roocode,
    author       = {{RooCodeInc}},
    title        = {{Roo-Code}},
    note         = {GitHub repository},
    url          = {https://github.com/RooCodeInc/Roo-Code},
    urldate      = {2025-11-17}
}

@Article{LAMMPS,
  author = "A. P. Thompson and H. M. Aktulga and R. Berger and 
     D. S. Bolintineanu and W. M. Brown and P. S. Crozier and
     P. J. in 't Veld and A. Kohlmeyer and S. G. Moore and T. D. Nguyen and
     R. Shan and M. J. Stevens and J. Tranchida and C. Trott and S. J. Plimpton",
  title = "{LAMMPS} - a flexible simulation tool for
     particle-based materials modeling at the 
     atomic, meso, and continuum scales",
  journal = "Comp. Phys. Comm.",
  volume =  "271",
  pages =   "108171",
  year =    "2022",
  doi = "10.1016/j.cpc.2021.108171"
}

@article{qmcpack,
  title={QMCPACK: Advances in the development, efficiency, and application of auxiliary field and real-space variational and diffusion quantum Monte Carlo},
  author={Kent, Paul RC and Annaberdiyev, Abdulgani and Benali, Anouar and Bennett, M Chandler and Landinez Borda, Edgar Josu{\'e} and Doak, Peter and Hao, Hongxia and Jordan, Kenneth D and Krogel, Jaron T and Kyl{\"a}np{\"a}{\"a}, Ilkka and others},
  journal={The Journal of chemical physics},
  volume={152},
  number={17},
  year={2020},
  publisher={AIP Publishing}
}

@inproceedings{hlspilot,
  title={Hlspilot: Llm-based high-level synthesis},
  author={Xiong, Chenwei and Liu, Cheng and Li, Huawei and Li, Xiaowei},
  booktitle={Proceedings of the 43rd IEEE/ACM International Conference on Computer-Aided Design},
  pages={1--9},
  year={2024}
}

@article{hlsrewriter,
  title={HLSRewriter: Efficient Refactoring and Optimization of C/C++ Code with LLMs for High-Level Synthesis},
  author={Xu, Kangwei and Zhang, Grace Li and Yin, Xunzhao and Zhuo, Cheng and Schlichtmann, Ulf and Li, Bing},
  journal={ACM Transactions on Design Automation of Electronic Systems},
  year={2025},
  publisher={ACM New York, NY}
}

@article{lift,
  title={LIFT: Llm-based pragma insertion for HLS via GNN supervised fine-tuning},
  author={Prakriya, Neha and Ding, Zijian and Sun, Yizhou and Cong, Jason},
  journal={arXiv preprint arXiv:2504.21187},
  year={2025}
}

@article{chathls,
  title={ChatHLS: Towards Systematic Design Automation and Optimization for High-Level Synthesis},
  author={Li, Runkai and Xiong, Jia and He, Xiuyuan and Lv, Jiaqi and Zhao, Jieru and Wang, Xi},
  journal={arXiv preprint arXiv:2507.00642},
  year={2025}
}

@article{llm-dse,
  title={LLM-DSE: Searching Accelerator Parameters with LLM Agents},
  author={Wang, Hanyu and Wu, Xinrui and Ding, Zijian and Zheng, Su and Wang, Chengyue and Nowatzki, Tony and Sun, Yizhou and Cong, Jason},
  journal={arXiv preprint arXiv:2505.12188},
  year={2025}
}

@inproceedings{hls-repair,
  title={Automated c/c++ program repair for high-level synthesis via large language models},
  author={Xu, Kangwei and Zhang, Grace Li and Yin, Xunzhao and Zhuo, Cheng and Schlichtmann, Ulf and Li, Bing},
  booktitle={Proceedings of the 2024 ACM/IEEE International Symposium on Machine Learning for CAD},
  pages={1--9},
  year={2024}
}

@article{c2hlsc,
  title={C2hlsc: Leveraging large language models to bridge the software-to-hardware design gap},
  author={Collini, Luca and Garg, Siddharth and Karri, Ramesh},
  journal={ACM Transactions on Design Automation of Electronic Systems},
  volume={30},
  number={6},
  pages={1--24},
  year={2025},
  publisher={ACM New York, NY}
}

@article{sage-hls,
  title={SAGE-HLS: Syntax-Aware AST-Guided LLM for High-Level Synthesis Code Generation},
  author={Khan, M and Mashnoor, Nowfel and Akyash, Mohammad and Azar, Kimia and Kamali, Hadi},
  journal={arXiv preprint arXiv:2508.03558},
  year={2025}
}

\end{document}